\documentclass[journal]{IEEEtran}

\ifCLASSINFOpdf
\else
   \usepackage[dvips]{graphicx}
\fi
\usepackage{url}

\usepackage{mathrsfs}
\usepackage{graphicx}
\usepackage{cite}
\usepackage{amssymb}
\usepackage{amsmath}
\usepackage{booktabs}
\usepackage{multirow}
\usepackage{pifont} 
\usepackage{hyperref}  
\hypersetup{
	colorlinks=false,   
	linkcolor=red,    
	urlcolor=red      
}

\usepackage[table]{xcolor}

\definecolor{ANNRow}{RGB}{245,248,255}  
\definecolor{BNNRow}{RGB}{255,248,238}  

\renewcommand{\baselinestretch}{0.95}
\newcommand{\cmark}{\ding{51}}
\newcommand{\xmark}{\ding{55}}
\newcommand{\lad}[1]{\textcolor{black}{#1}}

\begin{document}

\title{High-Fidelity Binarized Neural Vocoder}

\author{Andong Li,~\IEEEmembership{Member, IEEE}, Tong Lei, Xiaodong Li, Chengshi Zheng,~\IEEEmembership{Senior Member, IEEE}
\thanks{Andong Li, Xiaodong Li, Chengshi Zheng are with the Institute of Acoustics, Chinese Academy of Sciences, Beijing 100190, China, and are also with University of Chinese Academy of Sciences, Beijing 10190, China (e-mail:\{liandong, lxd, cszheng\}@mail.ioa.ac.cn). Tong Lei is with Tencent Hunyuan Multimodal Models Department, Tencent, Shenzhen 518054, China (email: fayelei@tencent.com). Chengshi Zheng is the corresponding author. This work was supported by the National Natural Science Foundation of China (NSFC) under Grants 62501588 and Young Elite Scientists Sponsorship Program of the Beijing High Innovation Plan.}
}

\maketitle

\begin{abstract}
 Neural vocoders have achieved remarkable progress in high-fidelity waveform generation in recent years, but most existing methods still rely on full-precision operators, leading to non-negligible storage and computational costs during inference. Binary neural networks (BNNs) provide an alternative direction by quantizing weights and activations to 1 bit. However, directly applying binarization to neural vocoders is rather challenging as it inherently incurs severe information loss. To this end, we propose BiVoC, a BNN-oriented time-frequency neural vocoder for high-fidelity waveform generation. Building on our previous full-precision RNDVoC framework, BiVoC adapts the inverse-problem formulation, range-null decomposition, and fine-to-coarse non-uniform subband modeling to the 1-bit setting. Specifically, we develop a simple binarized convolutional unit to provide stable time-frequency feature modeling under binary representations. Furthermore, we analyze the quality-complexity sensitivity of different modules and propose a mixed-precision design to retain quality-sensitive components in full precision while binarizing computation-dominant parts. Extensive experiments are conducted on LibriTTS and multiple out-of-distribution benchmarks. Both objective and subjective results demonstrate that BiVoC substantially outperforms BNN variants obtained by directly binarizing existing vocoders, while achieving competitive performance against several mainstream full-precision vocoder baselines under much lower theoretical computational cost. These results demonstrate the promising potential of BNNs in the audio generation field. Codes and pre-trained weights are available at https://github.com/Andong-Li-speech/BiVoC.
\end{abstract}


\IEEEpeerreviewmaketitle
\section{Introduction}
\IEEEPARstart{N}{eural} audio vocoders aim to synthesize time-domain waveforms from compact acoustic representations, such as mel-spectrograms or latent acoustic features~{\cite{siuzdakvocos}}. As a fundamental technique in the audio community, neural vocoders have been widely employed in various audio tasks, such as speech enhancement~{\cite{li2021two,shao2025cleanmel}}, audio codec~{\cite{li2026nanocodec,zeghidour2021soundstream}}, text-to-speech (TTS)~{\cite{tan2024naturalspeech}}, and multimodal generation~{\cite{liang2024dancecomposer}}. 

\begin{figure}[t]
	\centering
	\includegraphics[width=0.99\columnwidth]{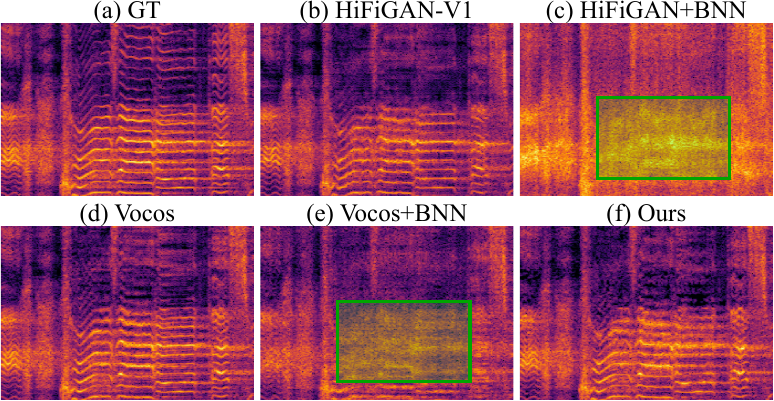}
	\caption{A typical spectral comparison, where the audio file is from MUSDB18 test set~{\cite{rafii2017musdb18}}. (a) Ground truth; (b) HiFiGAN-V1; (c) HiFiGAN+BNN, where float-based convolutional layers in HiFiGAN-V1 are replaced by the binarized counterparts; (d) Vocos; (e) Vocos+BNN, where float-based convolutional layers in Vocos are replaced by the binarized counterparts. (f) Proposed BiVoC.
	}
	\label{fig:demo-comparison}
	\vspace{-0.40cm}
\end{figure}

Traditional vocoders usually rely on manually extracted features, such as pitch and spectral envelope, for waveform synthesis. Seminal works include STRAIGHT~{\cite{kawahara2006straight}} and WORLD~{\cite{morise2016world}}, which, however, often suffer from audible distortion and unsatisfactory subjective quality. Recent advances in deep neural networks (DNNs) have substantially improved the reconstruction quality of neural vocoders, which have gradually become the mainstream approach. According to the input acoustic representations, existing methods can be roughly divided into two categories: mel-spectrogram and latent feature. The former adopts a fixed linear mel-filter to obtain compact time-frequency (T-F) representations, while the latter usually extracts acoustic features through stacked learnable nonlinear layers. In this paper, we focus on the former due to its wide applicability{\footnote{With the rapid development of generative models, neural audio codecs and variational autoencoders (VAEs) have attracted increasing attention for feature compression and reconstruction. Nonetheless, mel-spectrograms remain one of the dominant acoustic representations, especially when the compression ratio is high and the frame rate is relatively low~{\cite{wang2026optimising}}.}}. The investigation of latent-feature-based vocoders is left as future work.

For modern neural vocoder designs, processing efficiency and generation quality are two crucial factors. Early works are usually based on autoregressive~{\cite{van2016wavenet,kalchbrenner2018efficient}} or flow-based paradigms~{\cite{prenger2019waveglow,ping2020waveflow}}. Despite their notable performance improvements, these methods often suffer from low inference speed or prohibitive computational complexity. Subsequently, generative adversarial networks (GANs) have become an appealing solution due to their superiority in both generation quality and efficiency. Representative works include MelGAN~{\cite{kumar2019melgan}} and HiFiGAN~{\cite{kong2020hifi}}, where the mel-spectrogram is progressively expanded into waveform samples through stacked upsampling operations, and multi-perspective discriminators are adopted to detect generation artifacts across different modeling scales. Beyond waveform-domain generation, several recent studies propose to model the waveform from the T-F domain~{\cite{siuzdakvocos,li2025learning,li2025bridgevoc}}, that is, the generator estimates the spectral magnitude and phase component, followed by inverse short-time Fourier transform (iSTFT) for waveform synthesis. Compared with waveform-domain methods, T-F domain vocoders can achieve comparable or even better reconstruction quality with faster generation speed.

Nonetheless, the efficiency issue of neural vocoders is far from being fully addressed. Most existing high-fidelity vocoders are still built upon full-precision layers, where model storage, memory access, and convolutional computation remain non-negligible during inference. This issue becomes more critical in practical scenarios such as on-device TTS, where the vocoder is expected to run continuously under strict latency, power, and memory budgets. Such constraints are common on affordable, resource-constrained edge platforms, including low-cost service robots with spoken interaction, embedded human–machine interfaces in household or industrial equipment, and portable assistive or medical acoustic devices. These applications often require local speech generation or reconstruction to reduce response latency, tolerate intermittent connectivity, and protect user privacy, while operating with limited memory and energy budgets. This practical demand motivates the development of compact vocoders with reduced model storage and high-precision computation.

Existing approaches to efficient neural inference generally follow two complementary directions. The first is to reduce structural redundancy through compact architecture design or model pruning. Pruning typically identifies and removes redundant weights, filters, or channels from a pre-trained model, followed by fine-tuning to recover performance. Unstructured pruning provides flexible fine-grained sparsity, whereas structured pruning removes complete filters or channels and is generally more amenable to acceleration on general-purpose hardware~{\cite{han2015learning,li2017pruning}}. The second direction is to reduce the numerical precision of the remaining weights and activations through quantization. These two directions operate on different dimensions of model compression and can be jointly applied to obtain additional reductions in storage and computation~{\cite{han2015deep}}. In this study, we focus on the latter and investigate its extreme 1-bit form, which remains largely unexplored for high-fidelity neural vocoders.

Binary neural networks (BNNs) provide a promising path to this problem. By constraining weights and activations to binary values, BNNs can replace most floating-point multiplications with bitwise operations, leading to compact and hardware-friendly models~{\cite{hubara2016binarized}}. In theory, compared with a conventional convolutional layer, binary operations can achieve a 32$\times$ memory compression ratio and up to 64$\times$ computational reductions under suitable hardware support~{\cite{rastegari2016xnor}}. In recent years, BNNs have been explored in various image-related tasks, such as classification~{\cite{liu2018bi}} and restoration~{\cite{cai2023binarized}}. To the best of our knowledge, BNNs have not been systematically investigated for neural vocoding.

Directly binarizing neural vocoders remains a technically challenging task. Unlike image classification, neural vocoding requires the model to synthesize continuous waveforms with accurate temporal dependency, harmonic continuity, and phase structure. Small spectral errors can be directly perceived as audible artifacts. As shown in Fig.~{\ref{fig:demo-comparison}}(b)-(e), simply replacing the full-precision convolutional layers in existing neural vocoders, such as HiFiGAN and Vocos, with binarized counterparts leads to severe spectral degradation. This indicates that existing vocoder architectures cannot be binarized through na\"ive layer-wise replacement, mainly due to the excessive information loss caused by binary representations.

In~{\cite{li2025learning}}, we proposed RNDVoC, which introduced the formulation of mel-spectrogram vocoder as a linear inverse problem, the range-null decomposition framework, and the fine-to-coarse non-uniform subband modeling strategy. In this paper, we focus on a different question: \textit{given such a structured full-precision framework, how should its learnable modules, optimization strategy, and precision allocation be redesigned under the severe representation constraints imposed by 1-bit weights and activations?}

To answer this question, we propose \textbf{BiVoC}, a BNN-oriented T-F neural vocoder specifically designed for high-fidelity waveform reconstruction with 1-bit weights and activations. Instead of directly binarizing the full-precision modules, we re-engineer the learnable null-space estimator for binary computation. Specifically, the fine-to-coarse subband scheme is adapted to binary encoding and decoding, and we devise a simple binarized convolutional unit (BCU) with a stable residual feature flow. We further inspect the roles of different modules in the proposed framework and observe that preserving full-precision information in the encoder can notably improve reconstruction quality with only a slight increase in computational complexity. To the best of our knowledge, \textbf{BiVoC is the first BNN-oriented neural vocoder for high-fidelity waveform reconstruction} and achieves comparable generation quality to several mainstream full-precision methods including HiFiGAN-V1 and Vocos (see Fig.~{\ref{fig:demo-comparison}} for details), while requiring substantially lower computational cost.

The remainder of the paper is organized as follows. Sec.~{\ref{sec:related-work}} reviews the related work. Our proposed method is introduced in Sec.~{\ref{sec:proposed-method}}. Datasets and experimental setups are provided in Sec.~{\ref{sec:datasets}}, and we present extensive experiments in Sec.~{\ref{sec:results-analysis}}. Sec.~{\ref{sec:conclusion}} concludes this paper.   
\vspace{-0.25cm}
\section{Related Work}\label{sec:related-work}
\subsection{Neural Vocoder Methods}\label{sec:neural-vocoders}
\subsubsection{Autoregressive-based} WaveNet~{\cite{van2016wavenet}} formulates waveform generation as an autoregressive probabilistic modeling problem, where each waveform sample is predicted conditioned on previously generated samples, and multiple dilated convolution layers are adopted to gradually enlarge the receptive field. In~{\cite{valin2019lpcnet}}, the linear prediction coefficient (LPC) is adopted to estimate the current sample, followed by a lightweight RNN for residual prediction. Due to their autoregressive nature, these methods usually suffer from slow inference speed.
\subsubsection{Flow-based} Flow-based methods aim to establish a bijective mapping between simple Gaussian and target distributions. Representative works include WaveGlow~{\cite{prenger2019waveglow}} and FlowWaveNet~{\cite{kim2019flowavenet}}. Despite their improved generation speed over autoregressive-based vocoders, they are usually constrained by carefully designed invertible blocks, and often involve higher training and inference costs.
\subsubsection{GAN-based} HiFiGAN~{\cite{kong2020hifi}} adopts multi-scale and multi-periodicity discriminators to improve the naturalness of generated waveforms. Later, BigVGAN~{\cite{leebigvgan}} introduces periodic activation function to provide a useful inductive bias for audio synthesis and further demonstrates the benefit of model scaling. Recently, several T-F domain vocoders have also been proposed. Vocos~{\cite{siuzdakvocos}} stacks ConvNext blocks and uses two separate output heads to estimate spectral magnitude and phase, respectively. DegVoC~{\cite{li2026degvoc}} regards vocoding from mel-spectrogram as a degradation-restoration problem and introduces a subband-based architecture, achieving near state-of-the-art (SoTA) performance among GAN-based methods. 
\subsubsection{Diffusion-based} With the rapid development of diffusion models, a number of diffusion-based vocoders have been proposed in recent years. DiffWave~{\cite{kongdiffwave}} is one of the earliest methods to incorporate denoising diffusion probabilistic models (DDPMs) into waveform generation. More recently, flow-matching (FM)~{\cite{lipmanflow}} has attracted increasing attention due to its simple formulation. Lee~\emph{et al.}~{\cite{lee2024periodwave}} utilized a multi-periodic estimator, and optimal transport FM (OTFM) is employed for trajectory modeling. Meanwhile, RFWave~{\cite{liu2025rfwave}} introduces a subband-based velocity estimator to facilitate more fine-grained spectral generation. 
\begin{figure*}[t]
	\centering
	\includegraphics[width=0.85\textwidth]{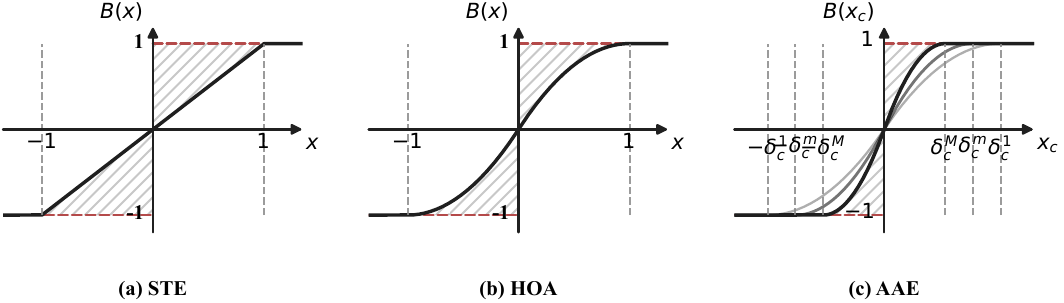}
	\caption{Comparisons among different gradient approximation strategies. (a) Straight through estimator; (b) High-order approximation estimator; (c)  Adaptive approximation estimator. 
	}
	\label{fig:gradient-comparison}
	\vspace{-0.40cm}
\end{figure*}
\vspace{-0.25cm}
\subsection{Binarization}\label{sec:binarization}
BNN is the extreme case of model quantization as it quantizes the weights and activations into only 1 bit, which significantly decreases the model size and computational complexity. In the vision field, it has been widely used in high-level~{\cite{hubara2016binarized,rastegari2016xnor,qin2020forward}} and low-level tasks~{\cite{cai2023binarized,zhang2024flexible}}. Beyond vision, BNNs have also been explored in speech and audio applications, including automatic speech recognition (ASR)~\cite{xiang2017binary,qian2019binary}, and speaker verification~\cite{zhu2021binary}. These studies mainly target classification or analysis-related problems. In contrast, neural vocoding requires sample-level waveform synthesis with accurate harmonic continuity and phase structure, making it more sensitive to the information loss caused by binarized weights and activations. Therefore, the use of BNNs for high-fidelity audio generation remains largely unexplored.

Another related line of research is spiking neural networks (SNNs), which process information through sparse spike events and have shown promising potential for ultra-low-power signal processing~{\cite{chen2025spiking,liu2026sse}}. However, SNNs usually require specialized designs such as spike encoding, neuron dynamics, and multi-step temporal simulation, rendering their training and deployment less straightforward than conventional ANN-based models. In contrast, BNNs preserve the standard feed-forward computation graph and can be more directly integrated into existing neural vocoder architectures, while still replacing most full-precision operations with bitwise computations.

\vspace{-0.25cm}
\section{Proposed Method}\label{sec:proposed-method}
\subsection{Preliminaries: Binary Neural Network}\label{sec:binary-neural-network}
In this subsection, we briefly introduce basic units for binary operations. 

\textbf{Forward Process}: Without loss of generality, for T-F domain scenario, we denote the full-precision two-dimensional (2-D) feature maps and kernels as $\mathbf{\mathcal{X}}^{f}\in\mathbb{R}^{C_{in}\times N\times T}$ and $\mathbf{\mathcal{W}}^{f}\in\mathbb{R}^{C_{out}\times C_{in}\times K_{n}\times K_{t}}$, where superscript $\left(\cdot\right)^{f}$ denotes the full-precision case, and $\left\{C, N, T\right\}$ are channel, frequency and frame sizes, respectively. Similarly, let $\mathcal{X}^{b}$ and $\mathcal{W}^{b}$ denote the binarized feature maps and kernels, respectively, and superscript $\left(\cdot\right)^{b}$ denotes the binary case. Here, $\mathbb{R}$ describes the algorithmic real-valued domain rather than prescribing FP32 as the mandatory hardware data type. Although the mel-spectrogram input is continuous-valued, the input and non-binary intermediate activations can be quantized into INT8 or another fixed-point format for integer deployment, while binary operands can be stored in a bit-packed 1-bit format. For an activation tensor with $N_{a}$ elements, FP32, INT8, and bit-packed binary representations require approximately $4N_{a}$, $N_{a}$, and $\left\lceil N_a / 8 \right\rceil$ bytes, respectively. The actual peak memory additionally depends on accumulators, residual branches, and operator workspaces. Our current PyTorch implementation uses floating-point tensors, while end-to-end integer deployment requires dedicated quantized kernels and is left for future work.

\begin{figure}[t]
	\centering
	\includegraphics[width=0.99\columnwidth]{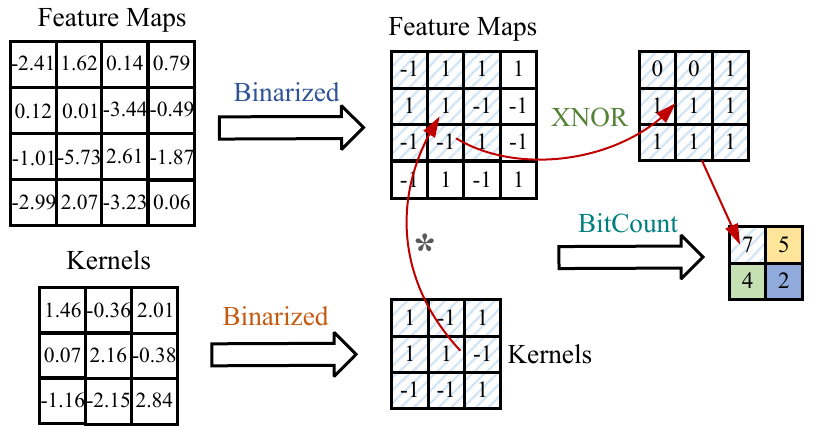}
	\caption{Illustration of binary convolution operation. Both $C_{in}$ and $C_{out}$ are set to 1. We ignore the scalar value and $\alpha_{c}$ is set to 0 for simplification.
	}
	\label{fig:example-bnn}
	\vspace{-0.40cm}
\end{figure}

For binary convolution, both weights and feature maps are binarized to -1 and +1, and efficient bitwise XNOR and bit-counting operations can then replace heavy floating-point matrix multiplication:
\begin{equation}
	\label{eqn:1}
	\mathcal{X}^{b}\otimes\mathcal{W}^{b} = \text{BitCount}\left(\text{XNOR}\left(\mathcal{X}^{b},\mathcal{W}^{b}\right)\right),
\end{equation}
\begin{equation}
	\label{eqn:2}
	x_{c,n,t}^b = \text{Sign}\left(x_{c,n,t}^f - \alpha_{c}\right) = 
	\begin{cases}
		+1, & \text{if } x_{c,n,t}^f > \alpha_{c} \\
		-1, & \text{if } x_{c,n,t}^f \leq \alpha_{c}
	\end{cases},
\end{equation}
\begin{align}
	\label{eqn:3}
	w_{i,j,k_{n},k_{t}}^b &= \frac{\left\| \mathcal{W}_{i,:}^f \right\|_1}{\mathcal{N}} \, \text{Sign}\left(w_{i,j,k_{n},k_{t}}^f\right) \\
	&= 
	\begin{cases}
		+\dfrac{\left\| \mathcal{W}_{i,:}^f \right\|_1}{\mathcal{N}}, & \text{if } w_{i,j,k_{n},k_{t}}^f > 0 \\[1.5em]
		-\dfrac{\left\| \mathcal{W}_{i,:}^f \right\|_1}{\mathcal{N}}, & \text{if } w_{i,j,k_{n},k_{t}}^f \leq 0
	\end{cases},
\end{align}
where $\left\{x_{c,n,t}^{b},x_{c,n,t}^{f}\right\}$ are elements of $\left\{\mathcal{X}^{b},\mathcal{X}^{f}\right\}$, with $c\in\left[1, C_{in}\right]$, $n\in\left[1, N\right]$, $t\in\left[1, T\right]$, $\left\{w_{i,j,k_{n},k_{t}}^{b}, w_{i,j,k_{n},k_{t}}^{f}\right\}$ are elements of $\left\{\mathcal{W}^{b}, \mathcal{W}^{f}\right\}$, with $i\in\left[1,C_{out}\right]$, $j\in\left[1, C_{in}\right]$, $k_{n}\in\left[1, K_{n}\right]$, $k_{t}\in\left[1, K_{t}\right]$. $\alpha_{c}$ is the threshold value for each input channel. Similar to~{\cite{rastegari2016xnor}}, $\frac{\left\| \mathcal{W}_{i,:}^f \right\|_1}{\mathcal{N}}$ denotes the average of absolute weight values, serving as a scalar to modulate the difference between binary and full-precision weights. We show an example of binary convolution in Fig.~{\ref{fig:example-bnn}}.

For the activation function, we adopt a ReAct-PReLU (RPReLU)~{\cite{liu2020reactnet}}, which is defined as:
\begin{equation}
	\label{eqn:5}
	f(y_{c,n,t}) = 
	\begin{cases}
		y_{c,n,t} - \gamma_{c} + \zeta_{c}, & \text{if } y_{c,n,t} > \gamma_{c} \\[1.5em]
		\beta_{c}\left(y_{c,n,t} - \gamma_{c}\right) + \zeta_{c}, & \text{if } y_{c,n,t} \leq \gamma_{c}
	\end{cases},
\end{equation}
where $y_{c,n,t}$ denotes the input feature. $\left\{\gamma_{c}, \zeta_{c}\right\}$ are learnable shifts to move the distribution, $\beta_{c}$ is a learnable parameter to control the slope of the negative part.

\textbf{Backward Propagation}: For quantization function in Eq.~{(\ref{eqn:2})}, the gradient calculation process can be formulated as:
\begin{equation}
	\label{eqn:6}
	\frac{\partial \mathcal{L}}{\partial x}
	=
	\frac{\partial \mathcal{L}}{\partial B(x)} \frac{\partial B(x)}{\partial x}
	=
	\frac{\partial \mathcal{L}}{\partial \mathrm{Sign}(x)} \frac{\partial \mathrm{Sign}(x)}{\partial x},
\end{equation} 
where $\mathcal{L}$ is the loss function, and $\mathcal{B}\left(x\right) = \text{Sign}\left(x\right)${\footnote{Here we ignore the subscript and superscript for simplification.}}. Notably, the derivative of the sign function is zero almost everywhere, and cannot be directly used during training. To this end, we introduce three surrogate functions for gradient estimation. 

The first type is the straight through estimator (STE)~{\cite{hubara2016binarized}}. As shown in Fig.~{\ref{fig:gradient-comparison}}(a), it is a clip function, given by:
\begin{equation}
	\label{eqn:7}
	\mathcal{S}\left(x_{c,n,t}^{f}\right) = \text{Clip}\left(x_{c,n,t}^{f}, -1, 1\right).
\end{equation}

The second type is the high-order approximation (HOA), as shown in Fig.~{\ref{fig:gradient-comparison}}(b). Compared with STE, HOA provides a smoother surrogate function and yields a more informative gradient within the clipping interval, given by:
\begin{equation}
	\label{eqn:8}
	\mathcal{S}(x_{c,n,t}^{f}) = 
	\begin{cases}
		-1, & x_{c,n,t}^{f} < -1 \\[1.2em]
		+(x_{c,n,t}^{f})^{2} + 2x_{c,n,t}^{f}, & -1 \leq x_{c,n,t}^{f} < 0 \\[1.2em]
		-(x_{c,n,t}^{f})^{2} + 2x_{c,n,t}^{f}, & 0 \leq x_{c,n,t}^{f} < 1 \\[1.2em]
		+1, & \text{otherwise}
	\end{cases},
\end{equation}
and the corresponding gradient can be represented as:
\begin{equation}
	\label{eqn:9}
	\mathcal{S}^{'}(x_{c,n,t}^{f}) = 
	\begin{cases}
		2x_{c,n,t}^{f} + 2, & -1 \leq x_{c,n,t}^{f} < 0 \\[1.2em]
		-2x_{c,n,t}^{f} + 2, & 0 \leq x_{c,n,t}^{f} < 1 \\[1.2em]
		0, & \text{otherwise}
	\end{cases}
\end{equation}

In~{\cite{xin2024rectified}}, an adaptive approximation estimator (AAE) was proposed, where learnable parameters are set for trade-off between clipping range and gradient representation accuracy, as shown in Fig.~{\ref{fig:gradient-comparison}}(c). Its surrogate function is defined as:
\begin{equation}
	\label{eqn:10}
	\mathcal{S}\left(x_{c,n,t}^{f}\right) = 
	\begin{cases}
		-1, & x_{c,n,t}^{f} < -\delta_{c} \\[1.2em]
		+\lambda_{c}^2 (x_{c,n,t}^{f})^{2} + 2\lambda_{c} x_{c,n,t}^{f}, & -\delta_{c} \leq x_{c,n,t}^{f} < 0 \\[1.2em]
		-\lambda_{c}^2 (x_{c,n,t}^{f})^{2} + 2\lambda_{c} x_{c,n,t}^{f}, & 0 \leq x_{c,n,t}^{f} < +\delta_{c} \\[1.2em]
		+1, & \text{otherwise}
	\end{cases},
\end{equation}
where $\lambda_{c}$ is a channel-wise learnable parameter, and the clip range $\delta_{c} = \frac{1}{\lambda_{c}}$. Note that HOA can be regarded as a special case of AAE when $\lambda_{c} = 1$.
\vspace{-0.25cm}
\subsection{Review of the Range-Null Decomposition}\label{sec:rnd}
BNN can incur severe information loss, and simply replacing a conventional convolution operation with its binary variant inevitably causes severe degradation in generation quality. Following RNDVoC~{\cite{li2025learning}}, we adopt the range-null decomposition (RND) formulation to reduce the reconstruction burden imposed on the binarized generator. For completeness, we briefly review its formulation below.

Without loss of generality, we consider a classical signal compression model:
\begin{equation}
	\label{eqn:11}
	\boldsymbol{y} = \mathbf{A}\boldsymbol{x} + \boldsymbol{n},
\end{equation}
where $\boldsymbol{y}\in\mathbb{R}^{d}$, $\boldsymbol{x}\in\mathbb{R}^{D}$, and $\boldsymbol{n}\in\mathbb{R}^d$ are observed, target, and noise signals, respectively. $\mathbf{A}\in\mathbb{R}^{d\times D}$ is a linear compression matrix with $d\ll D$. For the noise-free scenario, Eq.~{(\ref{eqn:11})} can be simplified into $\boldsymbol{y} = \mathbf{A}\boldsymbol{x}$. The solution of the equation  $\tilde{\boldsymbol{x}}$ should meet two constraints: (1) Degradation consistency: $\mathbf{A}\tilde{\boldsymbol{x}}\equiv\boldsymbol{y}$, which means the estimation should contain all the degradation information of the observation; (2) Distribution consistency: $\tilde{\boldsymbol{x}}\sim p\left(\boldsymbol{x}\right)$, which means that the estimation shares the same distribution as the target. Under these constraints, the solution can be written as:
\begin{equation}
	\label{eqn:12}
	\tilde{\boldsymbol{x}} = \mathbf{A}^{\dagger}\boldsymbol{y} + \left(\mathbf{I} - \mathbf{A}^{\dagger}\mathbf{A}\right)\hat{\boldsymbol{x}},
\end{equation}
where $\mathbf{A}^{\dagger}\in\mathbb{R}^{D\times d}$ is the pseudo-inverse of $\mathbf{A}$. Eq.~{(\ref{eqn:12})} consists of two terms, namely $\mathbf{A}^{\dagger}\boldsymbol{y}$ and $\left(\mathbf{I} - \mathbf{A}^{\dagger}\mathbf{A}\right)\hat{\boldsymbol{x}}$, which are called range-space (RS) and null-space (NS) components, respectively. Let $\mathbf{P}=\mathbf{A}^{\dagger}\mathbf{A}$. Since $\mathbf{P}$ is Hermitian and idempotent, it is an orthogonal projector, and thus
\begin{equation}
\label{eqn:13}
\left(\mathbf{A}^{\dagger}\boldsymbol{y}\right)^{\top}
\left(\mathbf{I}-\mathbf{A}^{\dagger}\mathbf{A}\right)\hat{\boldsymbol{x}}
=
\boldsymbol{x}^{\top}\mathbf{P}(\mathbf{I}-\mathbf{P})\hat{\boldsymbol{x}}=0.
\end{equation}

This suggests that the range-space component can be easily obtained by a fixed linear transform, and we only need to estimate the null-space part.

Recall that the mel-spectrogram can be formulated as:
\begin{equation}
	\label{eqn:14}
	\mathbf{Y} = \mathcal{A}\left|\mathbf{X}\right|,
\end{equation}
where $\mathbf{Y}\in\mathbb{R}^{N_m\times T}$ denotes the linear mel magnitude, obtained after inverse dynamic-range compression when log-mel features are used. Here, $N_m$ is the number of mel-frequency bins, $\mathcal{A}\in\mathbb{R}^{N_m\times N}$ is the mel-filter matrix, and $|\mathbf{X}|\in\mathbb{R}^{N\times T}$ denotes the target magnitude spectrum. 

In previous works, a neural network serves to estimate the target waveforms or T-F spectrogram given mel-spectrogram as the input, given by:
\begin{equation}
	\label{eqn:15}
	\boldsymbol{\tilde{x}} = \mathcal{F}_{time}\left(\mathbf{Y}\right)\quad\text{or}\quad \tilde{\mathbf{X}} = \mathcal{F}_{tf}\left(\mathbf{Y}\right),
\end{equation}
where $\mathcal{F}_{time}\left(\cdot\right)$ and $\mathcal{F}_{tf}\left(\cdot\right)$ are the corresponding mapping functions defined in the time and T-F domain, respectively. Notably, when BNN is adopted, the modeling capability can be dramatically hampered, leading to severe artifacts. The RND formulation alleviates this burden by separating the analytically recoverable range-space component from the learned null-space component.

Note that Eq.~{(\ref{eqn:14})} can be regarded as a special case of Eq.~{(\ref{eqn:11})}, where mel-filter $\mathcal{A}$ is an instantiation of the general compression matrix $\mathbf{A}$. Instead of following a direct-mapping paradigm for target estimation as in Eq.~{(\ref{eqn:15})}, similar to Eq.~{(\ref{eqn:12})}, we rewrite the solution as:
\begin{equation}
	\label{eqn:16}
	\left|\tilde{\mathbf{X}}\right| = \mathcal{A}^{\dagger}\mathbf{Y} + \left(\mathbf{I} - \mathcal{A}^{\dagger}\mathcal{A}\right)\mathcal{F}\left(\mathcal{A}^{\dagger}\mathbf{Y}\right).
\end{equation}

It involves two components. First, the mel-spectrogram is transformed into the linear scale via a linear operation, \emph{i.e.}, $\mathcal{A}^{\dagger}\mathbf{Y}$. After that, it acts as an initialization estimate and a network $\mathcal{F}\left(\cdot\right)$ is devised to estimate the remaining spectral part, filtered by $\left(\mathbf{I} - \mathcal{A}^{\dagger}\mathcal{A}\right)$. One may notice that only the magnitude is considered as the phase component is dropped in Eq.~{(\ref{eqn:14})}. Motivated by multi-task training, we enforce the network to also estimate the phase component, and Eq.~{(\ref{eqn:16})} is thus rewritten as:
\begin{equation}
	\label{eqn:17}
	\left\{\left|\hat{\mathbf{X}}\right|, \tilde{\mathbf{\Phi}}\right\} = \mathcal{F}\left(\mathcal{A}^{\dagger}\mathbf{Y}\right),
\end{equation}
\begin{equation}
	\label{eqn:18}
	\tilde{\mathbf{X}} = \left(\mathcal{A}^{\dagger}\mathbf{Y} + \left(\mathbf{I} - \mathcal{A}^{\dagger}\mathcal{A}\right)\left|\hat{\mathbf{X}}\right|\right)e^{j\tilde{\mathbf{\Phi}}},
\end{equation}
where $\left|\hat{\mathbf{X}}\right|$ denotes the residual magnitude, $\tilde{\mathbf{\Phi}}$ denotes the estimated phase. Compared with previous works, the rationales are two-fold. First, rather than modeling from scratch, it substantially mitigates the modeling burden by explicitly preserving the information from the mel observation, which is especially significant when the modeling capability is heavily restricted. Besides, as $\mathcal{A}^{\dagger}\mathbf{Y}$ has been aligned with the target spectrogram in the feature size after initialization, the subsequent reconstruction can be viewed as a T-F restoration problem, allowing us to borrow architectural principles from speech enhancement.
\begin{figure*}[t]
	\centering
	\includegraphics[width=0.98\textwidth]{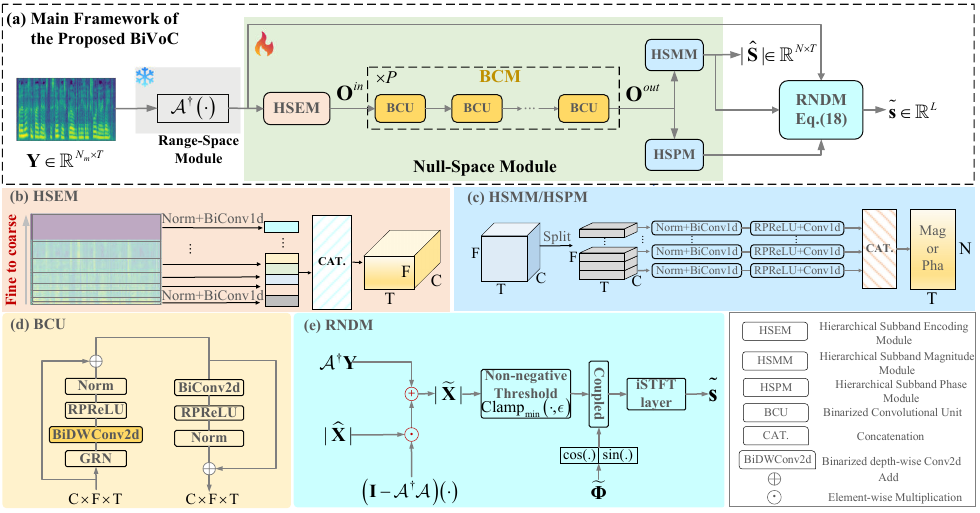}
	\caption{System diagram of the proposed BiVoC. (a) Overall structure of the BiVoC, which consists of three parts: RSM, NSM and RNDM; (b) Internal structure of the HSEM; (c) Internal structure of the HSMM/HSPM; (d) Internal structure of the BCU; (e) Internal structure of the RNDM.
	}
	\label{fig:system_diagram}
	\vspace{-0.40cm}
\end{figure*}
\vspace{-0.25cm}
\subsection{Proposed Framework}\label{sec:proposed-framework}
Based on the analysis in Sec.~{\ref{sec:rnd}}, in this paper, we develop a novel binary T-F vocoder called BiVoC based on range-null decomposition (RND) theory. The overall diagram is shown in Fig.~{\ref{fig:system_diagram}}(a). Following the modeling paradigm in~{\cite{li2025learning}}, the input mel-spectrogram is first converted into linear-scale counterpart via a range-space module (RSM). Then the null-space module attempts to estimate the residual magnitude spectrum as well as phase component, which consists of three components, namely hierarchical subband encoding module (HSEM), hierarchical subband magnitude/phase module (HSMM/HSPM), and binarized convolutional module (BCM). After that, the range-null decomposition module (RNDM) further generates the estimated waveform samples. The details are described below.
\subsubsection{HSEM} Previous T-F domain vocoders often perform spectrum generation with full-band modeling~{\cite{siuzdakvocos}}, \emph{i.e.}, the network is utilized to directly estimate the target spectrum, thus neglecting the hierarchical characteristics of subbands under different frequency regions. Besides, full-band modeling usually incurs excessive network parameters due to the use of large channel size. Recently, several subband-based modeling strategies have been proposed in the audio community~{\cite{li2025learning,li2025bridgevoc}}. To this end, we adopt a hierarchical strategy for spectrum encoding, as shown in Fig.~{\ref{fig:system_diagram}}(b). Specifically, given the input spectrum $\mathbf{Z}=\mathcal{A}^{\dagger}\mathbf{Y}\in\mathbb{R}^{N\times T}$, we explicitly split it into $F$ subbands in an uneven manner, \emph{i.e.}, $N = \sum_{i=1}^{F}N_{i}$, where $N_{i}$ denotes the frequency size of the $i$-th subband. As important acoustic cues, such as $F_{0}$ and harmonic structures, lie in the low- or mid-frequency regions, we follow a ``from-fine-to-coarse'' strategy, that is, we gradually increase $N_{i}$ as the subband index increases. For each subband, it is encoded with a separate normalization and binary Conv1d (BiConv1d)~{\cite{li2026nanocodec}}. After that, all encoded features are stacked into a 3-D tensor. The process is formulated as:
\begin{equation}
	\label{eqn:19}
	\mathbf{O}^{in}_{i} = \text{BiConv1d}\left(\text{Norm}\left(\mathbf{Z}_{b_{i-1}:b_{i}}\right)\right),
\end{equation}
\begin{equation}
	\label{eqn:20}
	\mathbf{O}^{in} = \text{CAT}\left(\mathbf{O}^{in}_{1},\cdots,\mathbf{O}^{in}_{F}\right)\in\mathbb{R}^{C\times F\times T},
\end{equation}
where $\left\{b_{i}=\sum_{j=1}^{i}N_j,\quad b_0=0\right\}$,
$C$ is the channel size, and $F$ denotes the number of subbands. Different from full-precision subband vocoders, BNN-based vocoders exhibit a distinct computational profile. Since binary convolution greatly reduces the cost of convolutional operations, full-precision components, especially normalization layers, account for a larger proportion of the overall complexity. Therefore, the normalization choice becomes significant for balancing reconstruction quality and inference efficiency. In Sec.~{\ref{sec:ablation-study}}, we compare the performance of different normalization types. 
\subsubsection{HSMM/HSPM} 
In contrast to HSEM, the decoding module attempts to estimate the magnitude and phase components, as shown in Fig.~{\ref{fig:system_diagram}}(c). Concretely, given the input $\mathbf{O}^{out}\in\mathbb{R}^{C\times F\times T}$, the $i$-th subband is passed through a normalization layer and BiConv1d, followed by a RPReLU activation. After that, a full-precision Conv1d is employed to estimate the corresponding magnitude or phase. The final layer is kept in full precision to produce continuous-valued spectral outputs, which is important for accurate magnitude and phase reconstruction. The process is given by:
\begin{equation}
	\label{eqn:21}
	\left|\hat{\mathbf{X}}\right|_{i} = \text{Exp}\left(\text{Conv1d}\left(\text{RPReLU}\left(\text{BiConv1d}\left(\text{Norm}\left(\mathbf{O}^{out}_{i}\right)\right)\right)\right)\right),
\end{equation} 
\begin{equation}
	\label{eqn:22}
	\mathbf{Q}_{i} = \text{Conv1d}\left(\text{RPReLU}\left(\text{BiConv1d}\left(\text{Norm}\left(\mathbf{O}^{out}_{i}\right)\right)\right)\right),
\end{equation}
\begin{equation}
	\label{eqn:23}
	\left\{\mathcal{R}\left(\mathbf{Q}_{i}\right), \mathcal{I}\left(\mathbf{Q}_{i}\right)\right\} = \text{Split}\left(\mathbf{Q}_{i}\right),
\end{equation}
\begin{equation}
	\label{eqn:24}
	\tilde{\mathbf{\Phi}}_{i} = \mathrm{Atan2}\left(\mathcal{I}\left(\mathbf{Q}_{i}\right), \mathcal{R}\left(\mathbf{Q}_{i}\right)\right),
\end{equation}
\begin{equation}
	\label{eqn:25}
	\left|\hat{\mathbf{X}}\right| = \text{CAT}\left(\left|\hat{\mathbf{X}}\right|_{1},\cdots,\left|\hat{\mathbf{X}}\right|_{F}\right)\in\mathbb{R}^{N\times T},
\end{equation}
\begin{equation}
	\label{eqn:26}
	\tilde{\mathbf{\Phi}} = \text{CAT}\left(\tilde{\mathbf{\Phi}}_{1},\cdots,\tilde{\mathbf{\Phi}}_{F}\right)\in\mathbb{R}^{N\times T},
\end{equation}
where $i\in\left\{1,\cdots,F\right\}$, $\mathcal{R}\left(\cdot\right)$ and $\mathcal{I}\left(\cdot\right)$ denote the real and imaginary parts.
$\text{Exp}\left(\cdot\right)$ and $\mathrm{Atan2}\left(\cdot\right)$ operations are adopted for magnitude and phase estimation, respectively.

\subsubsection{BCM} It is tasked with modeling deep features along the subbands and frame axes, which consists of $P$ binarized convolutional units (BCUs). Internal structure of the BCU is shown in Fig.~{\ref{fig:system_diagram}}(d), which includes two parts. The first part aims to aggregate contextual subband-frame information via a large kernel, and the second part is similar to feed-forward network (FFN) and facilitates information interaction along the channel dimension. Note that all convolution layers within BCU are binary versions. Specifically, for $p$-th unit, given the input $\mathbf{O}_{p-1}\in\mathbb{R}^{C\times F\times T}$, it first passes a global response normalization (GRN) layer~{\cite{woo2023convnext}}, then a depthwise BiConv2d (BiDWConv2d) is adopted for feature aggregation with the kernel size of $\left(K_{n}, K_{t}\right)$, followed by RPReLU and normalization with residual connection. The second part consists of a BiConv2d with group size of 2, RPReLU and another normalization layer. The whole process can be written as:
\begin{equation}
	\small
	\label{eqn:27}
	\mathbf{O}_{p}^{'} = \text{Norm}\left(\text{RPReLU}\left(\text{BiDWConv2d}\left(\text{GRN}\left(\mathbf{O}_{p-1}\right)\right)\right)\right) + \mathbf{O}_{p-1},
\end{equation}
\begin{equation}
	\label{eqn:28}
	\mathbf{O}_{p} = \text{Norm}\left(\text{RPReLU}\left(\text{BiConv2d}\left(\mathbf{O}_{p}^{'}\right)\right)\right) + \mathbf{O}_{p}^{'}.
\end{equation}
\subsubsection{RNDM} It aims to generate waveform samples from previous spectral estimation results, as shown in Fig.~{\ref{fig:system_diagram}}(e). Given the magnitude components from RSM and NSM, the target magnitude can be estimated following Eq.~{(\ref{eqn:18})}. Here $\left(\mathbf{I} - \mathcal{A}^{\dagger}\mathcal{A}\right)$ serves as a signal filter to preserve only the magnitude component defined in the null space. To guarantee the non-negativity of magnitude, a threshold is set, given by:
\begin{equation}
	\label{eqn:29}
	\left|\tilde{\mathbf{X}}\right| = \text{Clamp}_{\min}\left(\left|\mathbf{Z} + \left(\mathbf{I} - \mathcal{A}^{\dagger}\mathcal{A}\right)\left|\hat{\mathbf{X}}\right|\right|, \epsilon\right),
\end{equation} 
where $\epsilon$ is empirically set to 1e-5. After that, it is coupled with phase to yield the real and imaginary spectra, followed by an iSTFT layer to generate the waveform:
\begin{equation}
	\label{eqn:30}
	\tilde{\boldsymbol{s}} = \text{iSTFT}\left(\left|\tilde{\mathbf{X}}\right|\cos\left(\mathbf{\tilde{\Phi}}\right), \left|\tilde{\mathbf{X}}\right|\sin\left(\mathbf{\tilde{\Phi}}\right) \right).
\end{equation}
\textbf{Remarks}: One may ask \textit{why BiVoC is designed in this form and what makes it especially suitable for binarized neural vocoder}. Concretely, we answer the following three questions:\\
\noindent \ding{113}~(Q1): \textbf{Why use RND?}\\
\noindent \ding{113}~(Q2): \textbf{Why use hierarchical subband modeling?}\\
\noindent \ding{113}~(Q3): \textbf{Why design BCU in this way?}

For Q1, we argue that RND provides a principled way to reduce the modeling burden of the vocoder. Instead of enforcing a binary network to reconstruct the entire target spectrum, BiVoC obtains the range-space magnitude component through a fixed pseudo-inverse projection, and lets the network focus on estimating the remaining null-space magnitude component and phase. This decomposition mechanism is crucial for binarized neural vocoding, since the representation capacity of binary operations is inherently limited. In Sec.~{\ref{sec:ablation-study}}, we validate this point through ablation studies.

For Q2, we argue that hierarchical subband modeling is more compatible with the limited capacity of binary convolution. Full-band modeling forces the network to represent heterogeneous frequency regions in a unified space, which can easily lead to information loss and spectral distortion under binary operations. By contrast, subband modeling performs separate encoding for local frequency regions and conducts feature interaction across subbands and frames, allowing the model to capture region-specific spectral distributions with finer granularity. In Sec.~{\ref{sec:ablation-study}}, we validate the superiority of subband design through ablation studies.

For Q3, one seemingly straightforward solution is to replace the convolutional operations in existing powerful modules, such as ConvNeXtv2~{\cite{woo2023convnext}} blocks or LKCAB~{\cite{li2026degvoc}}, with binary counterparts. However, these modules usually contain complicated feature interaction mechanisms, including gating, feature modulation, or multiple nonlinear transformations. Under binary operations, such designs may amplify representation errors and disturb the spectral modeling process. We therefore follow a ``\textit{less is more}'' design philosophy and adopt an intentionally simple BCU. This simple residual design provides a more stable and effective feature flow for binarized neural vocoding, as validated in Sec.~{\ref{sec:ablation-study}}.
\begin{table}[t]
	\centering
	\caption{Parameter and OP of different NSM components under full-precision and binary settings. Following Eq.~(\ref{eqn:40}), $\mathrm{OPs}^{f}$ and $\mathrm{OPs}^{b}$ denote the computational operations under full-precision and binary settings, respectively. Values in parentheses denote the proportion of each component within the corresponding NSM. Parameters are reported in millions (M), and OPs are reported in Giga operations per 5-second audio clip. Percentages are calculated from unrounded values.}
	\large
	\label{tbl:nsm-complexity-distribution}
	\renewcommand{\arraystretch}{1.12}
	\setlength{\tabcolsep}{4.5pt}
	\resizebox{0.99\columnwidth}{!}{
		\begin{tabular}{@{}clccc@{}}
			\toprule
			Precision &Metric &HSEM &BCM &HSMM+HSPM \\
			\midrule
			\multirow{2}{*}{Full-precision}
			& \#Params (M) &0.28(4.33\%) &2.43(37.96\%) &3.70(57.71\%) \\
			&$\mathrm{OPs}^{f}$ (Giga/5s) &0.13(0.44\%) &28.70(93.78\%) &1.77(5.78\%) \\
			\midrule
			\multirow{2}{*}{Binary}
			&\#Params (M) &0.01(1.49\%) &0.20(26.27\%) &0.56(72.24\%) \\
			&$\mathrm{OPs}^{b}$ (Giga/5s) &0.006(0.54\%) &0.83(75.87\%) &0.26(23.59\%) \\
			\bottomrule
		\end{tabular}
	}
	\vspace{-0.25cm}
\end{table}
\begin{table}[t]
	\centering
	\caption{Fine-grained breakdown of the theoretical weight-storage and peak activation-buffer requirements of different NSM components. ``FP32'' denotes the full-precision counterpart, while ``Packed'' denotes the bit-packed binary implementation.}
	\label{tbl:nsm-memory-breakdown}
	\large
	\renewcommand{\arraystretch}{1.12}
	\setlength{\tabcolsep}{2.5pt}
	\resizebox{\columnwidth}{!}{
		\begin{tabular}{@{}llcccccc@{}}
		\toprule
		& & \multicolumn{3}{c}{Weight storage} 
		& \multicolumn{3}{c}{Peak activation I/O buffer} \\
		\cmidrule(lr){3-5}\cmidrule(lr){6-8}
		Component &Main operations
		&FP32 &Packed & Red.
		&FP32 &Packed &Red. \\
		& &(MB) &(MB) &(\%)
		&(MB) &(MB) &(\%) \\
		\midrule
		HSEM
		&Norm. + subband enc.
		&1.06 &0.07 &93.40
		&0.67 &0.47 &29.85 \\
		
		BCM
		&$8\times$(MS Conv + MLP)
		&9.30 &0.81 &91.29
		&22.95 &11.83 &48.45 \\
		
		HSMM+HSPM
		&Mag./phase dec.
		&14.11 &2.19 &84.46
		&0.92 &0.47 &48.91\\
		\bottomrule
		\end{tabular}
	}
	\vspace{-0.25cm}
\end{table}
\vspace{-0.25cm}
\subsection{Hybrid Design}\label{sec:hybrid-design}
As stated in Sec.~{\ref{sec:proposed-framework}}, except for the last projection layer in the HSMM/HSPM, all convolutional layers in the NSM are based on binary operations. The three NSM components, namely HSEM (head), BCM (body), and HSMM/HSPM (tail), have quite different parameter and OP distributions. In Table~{\ref{tbl:nsm-complexity-distribution}}, we report the proportions of trainable parameters and computational operations (OPs) under full-precision and binary settings, respectively. In both cases, the body part dominates the computational cost, while the tail part accounts for most of the parameters.

Beyond the parameter and operation distributions reported in Table~{\ref{tbl:nsm-complexity-distribution}}, Table~\ref{tbl:nsm-memory-breakdown} further separates the resource savings into weight-storage and activation-buffer reductions. For the full-precision counterpart, all convolutional weights and their input/output feature maps are represented in FP32. For the theoretical bit-packed implementation, the binary convolutional weights are stored using one bit per element, together with one FP32 scaling factor for each output channel, while the remaining non-binary parameters are retained in FP32. For activation memory, only the outputs of the binary activation functions, which serve as the inputs to binary convolutions, are assumed to be bit-packed. The convolution outputs, normalization features, nonlinear transformation outputs, and residual feature maps remain non-binary. Therefore, the reported activation reduction is calculated from the peak input/output buffer of the binary convolutional layers, instead of assuming that all intermediate feature maps can be stored with one bit.

As shown in Table~\ref{tbl:nsm-memory-breakdown}, bit packing notably reduces the model weight storage and the peak activation I/O buffer. HSEM exhibits the highest relative weight-storage reduction of 93.40\%, whereas HSMM/HSPM contributes the largest absolute saving, reducing its weight storage from 14.11 to 2.19 MB, \emph{i.e.}, by 11.92 MB. BCM reduces its weight storage by 8.49 MB and, more importantly, determines the overall peak activation-buffer requirement, which is reduced from 22.95 to 11.83 MB. These values represent theoretical storage and buffering requirements under bit-packed deployment. Note that the current PyTorch implementation retains floating-point tensors and therefore does not directly realize these memory reductions.

This observation suggests that different modules may have varying impacts on reconstruction quality and inference efficiency. Therefore, we propose a mixed-precision modeling strategy, which provides a practical way to balance the two aspects: quality-sensitive modules can be retained in full precision, while computation-dominant modules can be binarized to reduce the overall cost. We validate this design in Sec.~{\ref{sec:ablation-study}}.
\vspace{-0.3cm}
\subsection{Loss Function}\label{sec:loss-function}
We incorporate both reconstruction and adversarial losses for training.\\
\textbf{Reconstruction Loss}: Following previous literature~{\cite{li2025learning,li2026degvoc}}, it consists of mel-loss, real and imaginary (RI) loss, log-amplitude loss, phase loss, and consistency loss. Details are given below.\\
(1) Mel Loss: We adopt a multi-resolution Mel loss, defined as the mean absolute error (MAE) between the estimated and target log-scale mel-spectrograms:
\begin{equation}
	\label{eqn:31}
	\mathcal{L}_{mel} = \frac{1}{I}\sum_{i=1}^{I}\frac{1}{TN_{m}^{\left(i\right)}}\sum_{n,t}\left\|
	\log\left(\tilde{\mathbf{Y}}_{n,t}^{\left(i\right)}\right) - \log\left(\mathbf{Y}_{n,t}^{\left(i\right)}\right) \right\|_{1},
\end{equation}
where superscript $\left(\cdot\right)^{\left(i\right)}$ denotes the $i$-th resolution index. Seven resolutions are set, with window length of $\left[32, 64, 128, 256, 512, 1024, 2048\right]$, and hop length set to window\_length / 4. The corresponding mel bins are $\left[5, 10, 20, 40, 80, 160, 320\right]$.\\
(2) RI Loss: It is defined as the MAE between estimated and target RI spectra:
\begin{equation}
	\label{eqn:32}
	\begin{split}
		\mathcal{L}_{ri} &=  \frac{1}{TN}\sum_{n,t}\left(\left\|\mathcal{R}(\mathbf{\tilde{X}}_{n,t}) - \mathcal{R}\left(\mathbf{X}_{n,t}\right)\right\|_{1} \right. \\
		& \left. +\left\|\mathcal{I}(\mathbf{\tilde{X}}_{n,t}) - \mathcal{I}\left(\mathbf{X}_{n,t}\right)\right\|_{1}\right).
	\end{split}
\end{equation}
(3) Log-amplitude Loss: It is defined as the mean-square error (MSE) between estimated and target log-scale magnitude spectra: 
\begin{equation}
	\label{eqn:33}
	\mathcal{L}_{a} = \frac{1}{TN}\sum_{n,t}\left\|\log\left(\left|\tilde{\mathbf{X}}_{n,t}\right|\right) - \log\left(\left|\mathbf{X}_{n,t}\right|\right)\right\|_{2}^{2}.
\end{equation}
(4) Phase Loss: Following~{\cite{gompsnr}}, we adopt a weighted version of the omnidirectional phase (WOP) loss, given by:
\begin{equation}
	\label{eqn:34}
	\mathcal{L}_{p} = \frac{1}{9TN}\sum_{i,n,t}\frac{\left|\mathbf{X}_{n,t}\right|\cdot f_{AW}\left(\nabla_{i}\mathbf{\Phi}_{n,t}-\nabla_{i}\mathbf{\tilde{\Phi}}_{n,t}\right)}{\max\left(\left|\mathbf{X}\right|\right)},
\end{equation}
where $i\in\left\{1,\cdots,9\right\}$,  $f_{AW}\left(\cdot\right)$ denotes the anti-warping function~{\cite{ai2024low}}, and $\nabla\mathbf{\Phi}$ denotes the omnidirectional phase, which can extract the differential relations between the center phase bin and its neighboring bins{\footnote{Interested readers may refer to~{\cite{gompsnr}} for detailed illustrations.}.\\
(5) Consistency Loss: It is defined as the MSE between estimated spectrum and its consistency counterpart $\mathbf{\hat{X}}_{n,t}$~{\cite{wisdom2019differentiable}}, defined as:
\begin{equation}
	\label{eqn:35}
	\begin{split}
		\mathcal{L}_{c} &=  \frac{1}{TN}\sum_{n,t}\left(\left\|\mathcal{R}(\mathbf{\tilde{X}}_{n,t}) - \mathcal{R}\left(\hat{\mathbf{X}}_{n,t}\right)\right\|_{2}^{2} \right.\\ 
		& \left. +\left\|\mathcal{I}(\mathbf{\tilde{X}}_{n,t}) - \mathcal{I}(\mathbf{\hat{X}}_{n,t})\right\|_{2}^{2}\right),
	\end{split}
\end{equation}
where $\hat{\mathbf{X}} = \text{STFT}\left(\text{iSTFT}\left(\tilde{\mathbf{X}}\right)\right)$. The overall reconstruction loss can be defined as:
\begin{equation}
	\label{eqn:36}
	\mathcal{L}_{recon} = \lambda_{mel}\mathcal{L}_{mel} + \lambda_{ri}\mathcal{L}_{ri} + \lambda_{a}\mathcal{L}_{a} + \lambda_{p}\mathcal{L}_{p} + \lambda_{c}\mathcal{L}_{c},
\end{equation}
where $\left\{\lambda_{mel}, \lambda_{ri}, \lambda_{a}, \lambda_{p}, \lambda_{c}\right\}$ are weighting hyper-parameters and empirically set to $\left\{45, 45, 45, 100, 45\right\}$, respectively. 

For Adversarial loss, we incorporate multi-periodic discriminator (MPD)~{\cite{kong2020hifi}} and multi-resolution STFT discriminator (MSTFTD)~{\cite{jang2021univnet}}. The hinge-GAN is adopted, and the adversarial loss for discriminator is defined as:
\begin{equation}
	\label{eqn:37}
	\mathcal{L}_{d} = \frac{1}{M} \sum_{m=1}^{M} \left[ \max\left(0, 1-D_m(\boldsymbol{s})\right) + \max\left(0, 1+D_m(\tilde{\boldsymbol{s}})\right) \right],
\end{equation}
where $M$ is the number of sub-discriminators. For generator, the adversarial loss is defined as:
\begin{equation}
	\label{eqn:38}
	\mathcal{L}_{g} = \frac{1}{M} \sum_{m=1}^{M} \max\left(0, 1 - D_m\left( \tilde{\boldsymbol{s}} \right)\right).
\end{equation}

Besides, the feature matching loss is also incorporated:
\begin{equation}
	\label{eqn:39}
	\mathcal{L}_{fm} = \frac{1}{LM} \sum_{l,m} \left\| g_l^m\left( \tilde{\boldsymbol{s}} \right) - g_l^m\left(\boldsymbol{s} \right) \right\|_{1},
\end{equation}
where $g_{l}^{m}\left(\cdot\right)$ is the $l$-th layer for the $m$-th sub-discriminator. The weighting hyper-parameters for adversarial loss are empirically set to 1.
\vspace{-0.15cm}
\section{Datasets and Experimental Setups}\label{sec:datasets-experimental}
\subsection{Datasets}\label{sec:datasets}
The LibriTTS benchmark~{\cite{zen19_interspeech}} is adopted for training, which covers diverse acoustic recording conditions. We use all clips from subsets \{\textit{train-clean-100}, \textit{train-clean-360}, \textit{train-other-500}\} as the training set, and \textit{val-full} subset is used for model validation. For evaluation, five test sets are employed: (1) LibriTTS: 208 clips from \{\textit{dev-clean}, \textit{dev-other}\} are chosen, which are common test sets for vocoder evaluations~{\cite{leebigvgan,li2025bridgevoc,li2026degvoc}}. (2) EARS~{\cite{richter2024ears}}: Compared with LibriTTS, it is an expressive anechoic English corpus and covers diverse emotions. We sample 200 clips for testing. (3) AISHELL3~{\cite{shi2020aishell}}: It is a Mandarin dataset, and we sample 200 clips for testing. (4) RFWave~{\cite{liu2025rfwave}}: The authors provide a unified test set with 900 clips, which covers speech, sound effects and music. (5) MUSDB18~{\cite{rafii2017musdb18}}: A music dataset including bass, drums, vocals, and other instruments. We only use its test set. All clips are resampled to 24 kHz.
\renewcommand\arraystretch{1.00}
\begin{table*}[t]
	\centering
	\large
	\caption{Descriptions of the full-precision vocoders.}
	\resizebox{0.98\textwidth}{!}{
		\begin{tabular}{ccccccc}
			\hline\hline
			Ids &Methods &Neural Type &Technique Type &Publish &Domain &Code Link\\
			\hline
			1 &HiFiGAN-V1~{\cite{kong2020hifi}} &Full-precision   &GAN  &NeurIPS 2020 &T &https://github.com/jik876/hifi-gan\\
			2&iSTFTNet-V1~{\cite{kaneko2022istftnet}} &Full-precision  &GAN  &ICASSP 2022 &T &https://github.com/rishikksh20/iSTFTNet-pytorch\\
			3 &Avocodo~{\cite{bak2023avocodo}} &Full-precision  &GAN &AAAI 2023  &T &https://github.com/ncsoft/avocodo\\
			4 &BigVGAN-base~{\cite{leebigvgan}} &Full-precision &GAN &ICLR 2023  &T &https://github.com/NVIDIA/BigVGAN\\
			5 &BigVGAN~{\cite{leebigvgan}} &Full-precision  &GAN &ICLR 2023  &T &https://github.com/NVIDIA/BigVGAN\\
			6 &APNet2~{\cite{du2023apnet2}} &Full-precision  &GAN &NCMMSC 2023 &T-F &https://github.com/redmist328/APNet2\\
			7 &Vocos~{\cite{siuzdakvocos}} &Full-precision &GAN  &ICLR 2024 &T-F &https://github.com/gemelo-ai/vocos\\
			8 &PriorGrad~{\cite{leepriorgrad}} &Full-precision &DDPM &ICLR 2022 &T &https://github.com/microsoft/NeuralSpeech/tree/master/PriorGrad-vocoder\\
			9 &RNDVoC~{\cite{li2025learning}} &Full-precision &GAN &IJCAI 2025 &T-F &http://github.com/Andong-Li-speech/RNDVoC \\
			10 &DegVoC~{\cite{li2026degvoc}} &Full-precision &GAN &AAAI 2026 &T-F &https://github.com/Andong-Li-speech/DegVoC \\
			\hline\hline
	\end{tabular}}
	\label{tbl:descriptions-baselines}
	\vspace{-6pt}
\end{table*}
\vspace{-0.25cm}
\subsection{Miscellaneous Configurations}\label{sec:mis-configurations}
\subsubsection{Mel setups} The window size is 1024, with hop length set to 256, and 1024-point FFT is adopted. 100 mel bins are utilized in total, and the lower and upper frequencies are set to 0 and 12 kHz, respectively.
\subsubsection{Network setups} Empirically, twenty-five subbands are set, including twelve 250 Hz bandwidth subbands for low frequency followed by eight 500 Hz subbands, four 1 kHz subbands and one remaining subband. In Sec.~{\ref{sec:performance-effect-subband}}, we compare the performance under different number of subbands. The channel size $C$ is set to 256 across the whole network. For BCM, $P=8$ BCUs are adopted for deep modeling, and the kernel size $\left(K_{n}, K_{t}\right)$ of BiDWConv2d is set to $\left(5, 5\right)$, empirically. Layer normalization is adopted if not specified otherwise.
\subsubsection{Training setups} The batch size is set to 16, and each audio clip is segmented into 64 frames (\emph{i.e.}, 16384 samples) for stable training. The AdamW optimizer is adopted, with $\left(\beta_{1}=0.8, \beta_{2}=0.99\right)$ and learning rate initialized at 2e-4. The model is trained with 1 NVIDIA A100 GPU, and the generator and discriminator are updated for 1 million steps, respectively. 
\subsubsection{Baseline setups} 
\textbf{Since no established BNN-based neural vocoder baselines are available}, we build two BNN variants from representative vocoders, \emph{i.e.}, HiFiGAN-V1~{\cite{kong2020hifi}} and Vocos~{\cite{siuzdakvocos}}. Specifically, we keep the input and output convolutional layers in full precision, while replacing the convolutional operations in the intermediate deep modeling modules with binarized counterparts. The resulting baselines are termed \textit{HiFiGAN+BNN} and \textit{Vocos+BNN}. Besides, we compare with another 10 advanced ANN-based baselines: HiFiGAN-V1, iSTFTNet-V1~{\cite{kaneko2022istftnet}}, Avocodo~{\cite{bak2023avocodo}}, BigVGAN-base, BigVGAN~{\cite{leebigvgan}}, APNet2~{\cite{du2023apnet2}}, Vocos, PriorGrad~{\cite{leepriorgrad}}, RNDVoC~{\cite{li2025learning}} and DegVoC~{\cite{li2026degvoc}}. Details are listed in Table~{\ref{tbl:descriptions-baselines}}.
\begin{figure*}[t]
	\centering
	\includegraphics[width=0.99\textwidth]{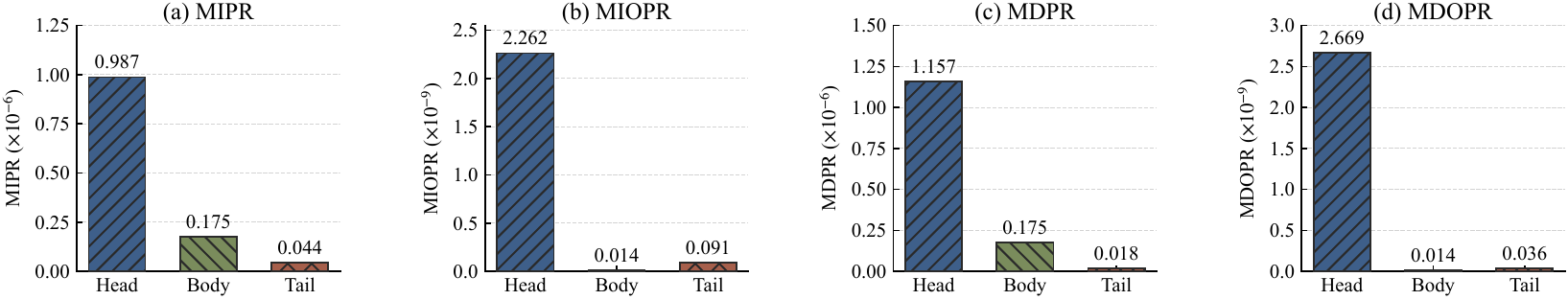}
	\caption{MIPR, MIOPR, MDPR, and MDOPR comparisons \emph{w.r.t} head, body, and tail components within the null-space module.}
	\label{fig:mipr-miopr}
	\vspace{-0.40cm}
\end{figure*} 
\vspace{-0.25cm}
\section{Results and Analysis}\label{sec:results-analysis}
\subsection{Evaluation Metrics}\label{sec:evaluation-metrics}
\subsubsection{Objective metrics} Eight metrics are involved: (1) Multi-resolution STFT (M-STFT)~{\cite{yamamoto2020parallel}}: which measures spectral distance across multiple STFT resolutions; (2) Wide-band perceptual evaluation of speech quality (PESQ)~{\cite{rec2005p}}, which quantifies the overall speech quality; (3) Mel-cepstral distortion (MCD)~{\cite{kubichek1993mel}}, where dynamic time warping (DTW) is utilized to calculate mel distance. (4) Periodicity RMSE and V/UV F$_{1}$ score~{\cite{morrisonchunked}}, which effectively measure the artifacts for non-autoregressive neural vocoders. (5) GOMPSNR~{\cite{gompsnr}}, an SNR-based metric, and is highly correlated with subjective quality by incorporating omnidirectional phase differential. (6) VISQOL~{\cite{hines2015visqol}}, an intrusive metric that scores spectro-temporal similarity between generated and reference signals. (7) Word error rate (WER), calculated with Whisper-Base by OpenAI.
\subsubsection{Subjective metrics} A total of twenty-one participants were recruited for the subjective listening test, including MOS testing and AB preference testing, and each participant was paid for 110 RMB. All participants self-reported normal hearing and had backgrounds in audio engineering. For the MOS test, each set includes fifteen reference-generation pairs, and participants were asked to rate six algorithms in each group on a [1, 5] scale (0.5 interval), according to the overall similarity to the reference. For the AB preference test, each set includes fifteen pairs, and participants were asked to select the clip with better overall quality in each pair. An ``Equal'' option was also provided when no clear preference could be made. The order of clips was randomly shuffled to mitigate rating bias. 
\subsubsection{Model Efficiency} We report the number of trainable parameters, computational operations (OPs) per five seconds, and inference speed to evaluate model efficiency. For full-precision models, all parameters and OPs are counted in the standard way. For BNN-based models, we follow the theoretical compression setting commonly used in BNN literature, where binary parameters are stored with 1 bit and binary operations are implemented by bitwise operations. Therefore, compared with their full-precision counterparts, binary parameters and OPs are counted as (1/32) and (1/64), respectively:
\begin{equation}
	\label{eqn:40}
	\mathrm{Params}^{b} = \frac{1}{32}\mathrm{Params}^{f},\quad \mathrm{OPs}^{b} = \frac{1}{64}\mathrm{OPs}^{f}.
\end{equation}

For inference speed on CPU and GPU, we evaluate on an Intel(R) Core(TM) i7-14700F and NVIDIA GeForce RTX 4060 Ti GPU, respectively{\footnote{Although BNNs can be accelerated by dedicated bitwise kernels or specialized hardware, our implementation is evaluated on general-purpose CPU/GPU platforms without hardware-specific XNOR-Popcount acceleration. Therefore, the measured inference speed should be regarded as a practical reference under common hardware settings, rather than the ideal theoretical acceleration of BNNs.}}.
\renewcommand\arraystretch{1.00}
\begin{table}[t]
	\centering
	\large
	\caption{Ablation studies on different surrogate functions for gradient. Layer normalization is adopted. Best results are highlighted in \textbf{BOLD}.}
	\label{tbl:surrogate-gradient}
	\resizebox{\linewidth}{!}{
		\begin{tabular}{c c c c c c}
			\toprule
			Type & M-STFT$\downarrow$ & PESQ$\uparrow$ & MCD$\downarrow$ & GOMPSNR$\uparrow$ & VISQOL$\uparrow$ \\
			\midrule
			STE  & 1.030 & 3.009 & 3.635 & 3.958 & 4.692 \\
			HOA  & 1.030 & 3.048 & 3.584 & 4.021 &4.699 \\
			AAE  & \textbf{1.024} & \textbf{3.055} & \textbf{3.568} & \textbf{4.045} & \textbf{4.705} \\
			\bottomrule
		\end{tabular}
	}
	\vspace{-0.25cm}
\end{table}
\renewcommand\arraystretch{1.00}
\begin{table*}[t]
	\centering
	\caption{Objective metric comparisons under different precision assignments for the head, body, and tail components of the null-space module. Layer normalization is adopted. ``M'' denotes a million parameters.}
	\label{tbl:bnn-module-ablation}
	\resizebox{0.90\textwidth}{!}{
		\begin{tabular}{ccccccccccc}
			\toprule
			\multirow{2}{*}{Ids} &\multicolumn{3}{c}{w/i BNN} &\#Params &OPs
			& \multirow{2}{*}{M-STFT$\downarrow$} 
			& \multirow{2}{*}{PESQ$\uparrow$} 
			& \multirow{2}{*}{MCD$\downarrow$} 
			& \multirow{2}{*}{GOMPSNR$\uparrow$}
			& \multirow{2}{*}{VISQOL$\uparrow$} \\
			\cmidrule(lr){2-4}
			 &Head & Body & Tail &(in M) & (Giga/5s)
			& &  &  &  &  \\
			\midrule
			1 & \cmark & \cmark & \cmark &\textbf{0.78} &\textbf{1.22} &1.024 &3.055 &3.568 &4.045 &4.705 \\
			\midrule
			2 &\xmark & \cmark & \cmark &1.08 &1.34 &0.975 &3.349 &3.342 &4.540 &4.759 \\
			3 &\cmark & \xmark & \cmark &3.04 &29.10 &0.952 &3.451 &3.299 &4.911 &4.771\\
			4 &\cmark & \cmark & \xmark &3.93 &2.73 &0.986 &3.193 &3.489 &4.199 &4.735 \\
			\midrule
			5 &\cmark &\xmark &\xmark &6.11 &30.61 &0.944 &3.473 &3.331 &4.864 &4.789\\
			6 &\xmark &\cmark &\xmark &4.15 &2.86 &0.947 &3.425 &3.276 &4.681 &4.769 \\ 
			7 &\xmark &\xmark & \cmark &3.30 &29.22 &0.899 &3.765 &2.975 &5.603 &4.831 \\
			\midrule
			8 &\xmark & \xmark & \xmark &6.41 &30.74 &\textbf{0.884} &\textbf{3.820} &\textbf{2.936} &\textbf{5.643} &\textbf{4.845} \\
			\bottomrule
		\end{tabular}
	}
	\vspace{-0.25cm}
\end{table*}
\renewcommand\arraystretch{1.00}
\begin{table}[t]
	\centering
	\large
	\caption{Ablation studies on whether to use the RND strategy. Layer normalization is adopted.}
	\label{tbl:rnd-ablation}
	\resizebox{\linewidth}{!}{
		\begin{tabular}{c c c c c c}
			\toprule
			w/i RND & M-STFT$\downarrow$ & PESQ$\uparrow$ & MCD$\downarrow$ & GOMPSNR$\uparrow$ & VISQOL$\uparrow$ \\
			\midrule
			\xmark  &1.512 &2.676 &4.968 &3.464 &4.536 \\
			\cmark  & \textbf{0.975} & \textbf{3.349} & \textbf{3.342} & \textbf{4.540} & \textbf{4.759} \\
			\bottomrule
		\end{tabular}
	}
	\vspace{-0.25cm}
\end{table}
\begin{table}[t]
	\centering
	\large
	\caption{Ablation studies on whether the subband-based modeling strategy is used. Layer normalization is adopted.}
	\label{tbl:subband-ablation}
	\resizebox{\linewidth}{!}{
		\begin{tabular}{c c c c c c}
			\toprule
			w/i Subband & M-STFT$\downarrow$ & PESQ$\uparrow$ & MCD$\downarrow$ & GOMPSNR$\uparrow$ & VISQOL$\uparrow$ \\
			\midrule
			\xmark  &1.049 &3.117 &3.670 &4.531 &4.687 \\
			\cmark  & \textbf{0.975} & \textbf{3.349} & \textbf{3.342} & \textbf{4.540} & \textbf{4.759} \\
			\bottomrule
		\end{tabular}
	}
	\vspace{-0.25cm}
\end{table}
\begin{figure}[t]
\centering
\includegraphics[width=0.97\columnwidth]{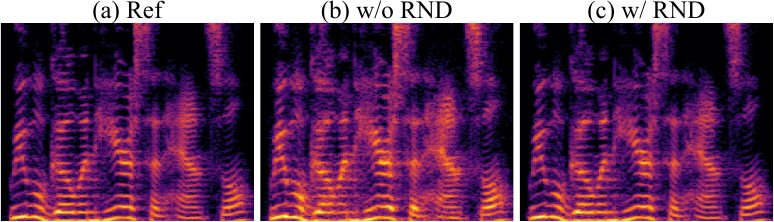}
\caption{A typical spectral comparison \emph{w.r.t} whether the RND strategy is utilized, where the audio file is from the LibriTTS test set.}
\label{fig:rnd-comparison}
\vspace{-0.35cm}
\end{figure}
\begin{table}[t]
	\centering
	\LARGE
	\caption{Ablation studies on different basic units. Layer normalization is adopted.}
	\label{tbl:units-ablation}
	\resizebox{\linewidth}{!}{
		\begin{tabular}{c c c c c c}
			\toprule
			Unit Type & M-STFT$\downarrow$ & PESQ$\uparrow$ & MCD$\downarrow$ & GOMPSNR$\uparrow$ & VISQOL$\uparrow$ \\
			\midrule
			ConvNextv2~{\cite{woo2023convnext}} &1.053 &2.890 &3.720 &3.842 &4.710 \\
			LKCAB~{\cite{li2026degvoc}} &1.005 &3.169 &3.494  &4.185 &4.743 \\
			BCU w/o Norm  &0.982 &3.287 & 3.433 &4.382 &\textbf{4.763} \\
			BCU (Ours)  &\textbf{0.975} &\textbf{3.349} &\textbf{3.342} &\textbf{4.540} &4.759 \\
			\bottomrule
		\end{tabular}
	}
	\vspace{-0.25cm}
\end{table}
\renewcommand\arraystretch{1.00}
\setcounter{table}{9}
\begin{table*}[t]
	\centering
	\large
	\caption{Objective comparisons among different baselines on the LibriTTS benchmark. ``$a\times$'' denotes the speed-up ratio over real-time. $\left(\cdot\right)^{\ddagger}$ indicates the use of pre-trained checkpoint from the original papers. The best performance within BNN category is highlighted with $\textbf{BOLD}$.}
	\resizebox{0.98\textwidth}{!}{
		\begin{tabular}{cccccccccccccc}
			\toprule
			\multirow{2}*{Models} &w/i &\#Params. &OPs &\multicolumn{2}{c}{Inference Speed} &\multirow{2}*{M-STFT$\downarrow$} &\multirow{2}*{PESQ$\uparrow$} &\multirow{2}*{MCD$\downarrow$} &V/UV$\uparrow$ &Period.$\downarrow$ &\multirow{2}*{GOMPSNR$\uparrow$}  &\multirow{2}*{VISQOL$\uparrow$} &WER$\downarrow$ \\
			\cmidrule{5-6}
			&BNN  &(M) &(Giga/5s) &CPU &GPU & & & &F1 &RMSE &  & &(in \%) \\
			\midrule
			GT &-  &- &- &- &- &- &- &- &- &- &- &- &6.21\\
			\rowcolor{ANNRow}
			HiFiGAN-V1 &\xmark &14.01 &166.41 &5.88$\times$ &157$\times$ &1.104 &3.056 &4.284 &0.921 &0.167 &3.293 &4.721 &7.89 \\
			\rowcolor{ANNRow}
			iSTFTNet-V1 &\xmark &13.33 &117.30 &10.66$\times$ &299$\times$ &1.157 &2.880 &4.480 &0.918 &0.167 &3.159 &4.655 &7.54 \\
			\rowcolor{ANNRow}
			Avocodo &\xmark &14.01 &166.47 &5.29$\times$ &148$\times$ &1.113 &3.217 &4.314 &0.913 &0.161 &3.541 &4.762 &7.25\\
			\rowcolor{ANNRow}
			BigVGAN-base &\xmark &14.01 &166.41 &2.33$\times$ &23$\times$ &0.884 &3.521 &3.034 &0.941 &0.131 &4.871 &4.869 &6.77 \\
			\rowcolor{ANNRow}
			BigVGAN &\xmark &112.39 &454.08 &1.54$\times$ &17$\times$ &0.812 &3.991 &2.547 &0.955 &0.104 &5.774 &4.934 &6.29 \\
			\rowcolor{ANNRow}
			APNet2 &\xmark &31.52 &14.79 &58.82$\times$ &749$\times$ &1.143 &2.834 &4.153 &0.923 &0.153 &4.015 &4.582 &7.44 \\
			\rowcolor{ANNRow}
			Vocos$^{\ddagger}$ &\xmark &13.53  &6.35 &142.76$\times$ &1538$\times$ &0.854 &3.615 &3.105 &0.948 &0.115 &5.005 &4.879 &6.61 \\
			\rowcolor{ANNRow}
			PriorGrad &\xmark &2.70  &8364 &0.56$\times$ &8$\times$ &1.393 &4.017 &5.854 &0.938 &0.130 &0.148 &4.737 &7.52 \\
			\rowcolor{ANNRow}
			RNDVoC &\xmark &3.14 &37.20 &15.17$\times$ &211$\times$ &0.746 &4.226 &1.914 &0.970 &0.074 &6.920 &4.915 &6.42 \\
			\rowcolor{ANNRow}
			DegVoC &\xmark &3.89 &45.62 &18.20$\times$ &217$\times$ &0.803 &4.225 &2.209 &0.966 &0.088 &6.812 &4.928 &6.29 \\
			\midrule
			\rowcolor{BNNRow}
			HiFiGAN+BNN &\cmark &\textbf{0.81} &2.80 &4.03$\times$ &121$\times$ &2.111 &1.246 &6.654 &0.673 &0.375 &1.065 &3.587 &16.18 \\
			\rowcolor{BNNRow}
			Vocos+BNN &\cmark &1.37 &\textbf{0.53} &\textbf{68.05$\times$} &\textbf{1062$\times$} &1.381 &1.829 &4.925 &0.869 &0.237 &2.041 &4.378 &9.46 \\
			\rowcolor{BNNRow}
			\textbf{BiVoC (Ours)} &\cmark &1.08 &1.01 &14.24$\times$ &191$\times$ &\textbf{0.962} &\textbf{3.447} &\textbf{3.222} &\textbf{0.938} &\textbf{0.134} &\textbf{4.645} &\textbf{4.779} &\textbf{6.98} \\
			\bottomrule
	\end{tabular}}
	\vspace{-0.3cm}
	\label{tbl:objective-metric-libritts}
\end{table*}
\renewcommand\arraystretch{1.00}
\setcounter{table}{8}
\begin{table}[t]
	\centering
	\large
	\caption{Ablation studies on different normalization types.}
	\label{tbl:norm-ablation}
	\resizebox{\linewidth}{!}{
		\begin{tabular}{c c c c c c}
			\toprule
			\multirow{2}{*}{Norm Type} &{\#Params} &OPs &\multirow{2}{*}{PESQ$\uparrow$} &\multirow{2}{*}{GOMPSNR$\uparrow$} & \multirow{2}{*}{VISQOL$\uparrow$} \\
			&(in M) &(Giga/5s) & & & \\
			\midrule
			LN &1.08 &1.34 &3.349 &4.540 &4.759 \\
			DyT &1.08 &1.12 &3.347 &4.512 &4.759  \\
			BN &1.08 &\textbf{1.01} &\textbf{3.447}  &\textbf{4.645} &\textbf{4.779} \\
			\bottomrule
		\end{tabular}
	}
	\vspace{-0.35cm}
\end{table}
\renewcommand\arraystretch{1.00}
\definecolor{FPBG}{HTML}{F3F6FA}
\definecolor{BNNBG}{HTML}{FFF6E8}
\newcolumntype{F}{>{\columncolor{FPBG}}c}
\newcolumntype{B}{>{\columncolor{BNNBG}}c}
\setcounter{table}{10}
\begin{table*}[t]
	\centering
	\LARGE
	\caption{Out-of-distribution objective comparisons among different vocoders on AISHELL3, EARS, and RFWave test sets. The best performance within BNN category is highlighted with \textbf{BOLD}.}
	\label{tbl:ood-objective-comparisons}
	\renewcommand{\arraystretch}{0.99}
	\resizebox{0.98\textwidth}{!}{
		\begin{tabular}{llFFFFFFFFFFBBB}
			\toprule
			\multirow{2}{*}{Datasets} 
			& \multirow{2}{*}{Metrics}
			& \multicolumn{10}{>{\columncolor{FPBG}}c}{Full-precision Vocoders}
			& \multicolumn{3}{>{\columncolor{BNNBG}}c}{BNN-based Vocoders} \\
			\cmidrule(lr){3-12} \cmidrule(lr){13-15}
			& 
			& HiFiGAN-V1
			& iSTFTNet-V1
			& Avocodo 
			& \shortstack{BigVGAN\\-base}
			& BigVGAN 
			& APNet2 
			& Vocos$^{\ddagger}$ 
			& PriorGrad 
			& RNDVoC
			& DegVoC
			& \shortstack{HiFiGAN\\+BNN}
			& \shortstack{Vocos\\+BNN}
			& \textbf{BiVoC (Ours)} \\
			\midrule
			
			\multirow{5}{*}{AISHELL3}
			& M-STFT$\downarrow$  &1.059 &1.132 &1.040 &0.858 &0.769 &0.975 &0.845 &1.054 &0.702 &0.747 &2.599 &1.355 &\textbf{0.897} \\
			& PESQ$\uparrow$ &2.823 &2.595 &3.003 &3.414 &3.878 &2.846 &3.342 &3.995 &4.049 &4.075 &1.167 &1.742 &\textbf{3.330} \\
			& MCD$\downarrow$ &2.827 &3.061 &2.864 &1.902 &1.602 &2.743 &2.046 &1.717 &1.267 &1.508 &6.327 &3.666 &\textbf{1.950} \\
			& GOMPSNR$\uparrow$  &3.351 &3.242 &3.721 &4.860 & 5.746 &4.182 &4.549 &6.067 &6.436 &6.560 &1.039 &2.423 & \textbf{4.928} \\
			& VISQOL$\uparrow$  &4.651 &4.486 &4.731 &4.806 &4.907 &4.715 &4.836 &4.675 &4.897 &4.904 &3.249 &4.331 &\textbf{4.737} \\
			\midrule
			\multirow{5}{*}{EARS}
			&M-STFT$\downarrow$ &1.210 &1.245 &1.197 &1.025 &0.839 &1.202 &0.898 &1.062 &0.754 &0.769 &2.613 &1.526 &\textbf{0.990} \\
			& PESQ$\uparrow$ &2.907 &2.604 &3.065 &3.492 &3.892 &2.450 &3.522 &3.847 &3.989 &4.147 &1.239 &1.783 & \textbf{3.163} \\
			& MCD$\downarrow$  &2.733 &2.819 &2.703 &1.871 &1.400 &2.768 &1.949 &1.692 &0.951 &1.114 &4.804 &3.193 &\textbf{1.979} \\
			& GOMPSNR$\uparrow$  &2.923 &2.583 &2.972 &4.227 & 5.287 &3.098 &4.495 &5.374 &6.307 &6.452 &0.636 &1.726 & \textbf{4.148} \\
			& VISQOL$\uparrow$ &4.570 &4.480 &4.641 &4.739 &4.898 &4.456 &4.848 &4.636 &4.869 &4.903 &3.085 &4.157 & \textbf{4.673} \\
			\midrule
			
			\multirow{5}{*}{RFWave}
			& M-STFT$\downarrow$  &1.263 &1.352 &1.280 &1.033 &0.933 &1.487 &0.950 &1.285 &0.793 &0.935 &3.664 &1.728 &\textbf{0.998} \\
			&PESQ$\uparrow$  &2.558 &2.342 &2.756 &3.080 &3.585 & 2.458 &3.186 &3.508 &3.839 &3.758 &1.198 &1.652 & \textbf{3.180} \\
			&MCD$\downarrow$   &4.703 &4.997 &4.634 &3.590 &3.081 &5.315 &3.408 &3.620 &2.470 &3.075 &9.698 &6.374 &\textbf{3.430} \\
			&GOMPSNR$\uparrow$    &2.699 &2.536 &2.958 &3.755 &4.720 &3.231 &3.880 &3.989 &5.681 &5.477 &0.658 &1.685 &\textbf{4.147} \\
			&VISQOL$\uparrow$   &4.509 &4.409 &4.590 &4.730 &4.842 &4.527 &4.826 &4.569 &4.907 &4.844 &2.968 &4.084 & \textbf{4.718} \\
			\bottomrule
		\end{tabular}
	}
	\vspace{-0.25cm}
\end{table*}
\begin{figure*}[h]
	\centering
	\includegraphics[width=0.98\textwidth]{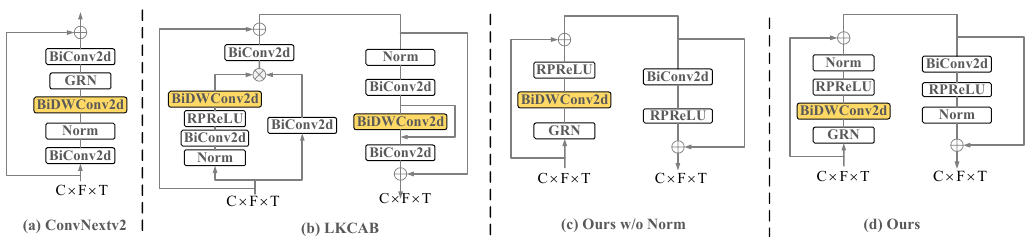}
	\caption{Internal structure of different basic modeling units. (a) ConvNextv2 block; (b) LKCAB block; (c) Our proposed BCU without normalization layer; (d) Our proposed BCU.
	}
	\label{fig:ablation-basic-units}
	\vspace{-0.40cm}
\end{figure*}
\vspace{-0.25cm}
\subsection{Ablation Studies}\label{sec:ablation-study}
We conduct extensive experiments on the LibriTTS test set.

\textbf{Different surrogate functions for gradient estimation}: We first compare different surrogate functions for gradient estimation. As shown in Table~{\ref{tbl:surrogate-gradient}}, STE yields the lowest overall objective performance among the three surrogate types, suggesting that the simple clipped surrogate is insufficient for optimizing high-fidelity binarized vocoders. HOA improves the results by providing a smoother high-order approximation, while AAE further achieves slightly better performance by adaptively adjusting the clipping range for different channels. This indicates that a more informative and distribution-aware surrogate gradient is important for mitigating the optimization difficulty of BNN-based vocoders. We thus choose AAE as the surrogate function for optimization by default.

\textbf{Impact of hybrid design}: To quantitatively evaluate the benefit of mixed-precision design, we analyze module-wise precision sensitivity from both the full binarized and full-precision baselines. Let $\mathcal{M}^{b}$ and $\mathcal{M}^{f}$ denote the full binarized and full-precision baselines, respectively. Further, let $\mathcal{M}^{m\rightarrow f}$ denote the hybrid variant in which module $m\in\{\mathrm{head}, \mathrm{body}, \mathrm{tail}\}$ is changed from binary to full precision while the other modules remain binary, and let $\mathcal{M}^{m\rightarrow b}$ denote the complementary variant in which only module $m$ is changed from full precision to binary. We then define four indicators: metric improvement parameter ratio (MIPR), metric improvement OP ratio (MIOPR), metric degradation parameter ratio (MDPR), and metric degradation OP ratio (MDOPR), as follows:

\begin{equation}
	\label{eqn:41}
	\operatorname{MIPR}^{m}=
	\frac{\operatorname{Metric}\left(\mathcal{M}^{m\rightarrow f}\right)
		-\operatorname{Metric}\left(\mathcal{M}^{b}\right)}
	{\operatorname{Params}\left(\mathcal{M}^{m\rightarrow f}\right)
		-\operatorname{Params}\left(\mathcal{M}^{b}\right)},
\end{equation}
\begin{equation}
	\label{eqn:42}
	\operatorname{MIOPR}^{m}=
	\frac{\operatorname{Metric}\left(\mathcal{M}^{m\rightarrow f}\right)
		-\operatorname{Metric}\left(\mathcal{M}^{b}\right)}
	{\operatorname{OPs}\left(\mathcal{M}^{m\rightarrow f}\right)
		-\operatorname{OPs}\left(\mathcal{M}^{b}\right)},
\end{equation}
\begin{equation}
	\label{eqn:43}
	\operatorname{MDPR}^{m}=
	\frac{\operatorname{Metric}\left(\mathcal{M}^{f}\right)
		-\operatorname{Metric}\left(\mathcal{M}^{m\rightarrow b}\right)}
	{\operatorname{Params}\left(\mathcal{M}^{f}\right)
		-\operatorname{Params}\left(\mathcal{M}^{m\rightarrow b}\right)},
\end{equation}
\begin{equation}
	\label{eqn:44}
	\operatorname{MDOPR}^{m}=
	\frac{\operatorname{Metric}\left(\mathcal{M}^{f}\right)
		-\operatorname{Metric}\left(\mathcal{M}^{m\rightarrow b}\right)}
	{\operatorname{OPs}\left(\mathcal{M}^{f}\right)
		-\operatorname{OPs}\left(\mathcal{M}^{m\rightarrow b}\right)}.
\end{equation}
where positive improvement for higher-is-better metrics and negative reduction for lower-is-better metrics. Notably, a larger MIPR or MIOPR indicates a larger performance per unit additional resource, while a larger MDPR or MDOPR indicates a larger performance degradation per unit resource saving.
\begin{table*}[t]
	\centering
	\caption{Subjective MOS among different vocoders on the out-of-distribution AISHELL3 and MUSDB18 benchmarks. We report mean with 95\% confidence intervals (CI95), and we perform a Friedman test to evaluate the overall statistical significance.}
	\resizebox{0.98\textwidth}{!}{
		\begin{tabular}{ccccccccc}
			\hline
			Models &GT &\lad{HiFiGAN-V1} &Vocos &APNet2 &HiFiGAN+BNN &Vocos+BNN &BiVoC (Ours) &\lad{Friedman $p$-value} \\
			AISHELL3 &4.32$\pm$0.08 &3.93$\pm$0.09 &\textbf{4.07$\pm$0.08} &3.69$\pm$0.09 &1.49$\pm$0.05 &2.64$\pm$0.08 &4.00$\pm$0.08 &$<0.001$\\
			MUSDB18 (Vocals) &4.22$\pm$0.08 &3.38$\pm$0.09    &3.49$\pm$0.08 &2.84$\pm$0.08 &1.35$\pm$0.04 &2.11$\pm$0.07 &\textbf{3.61$\pm$0.09} &$<0.001$ \\
			\hline
	\end{tabular}}
	\label{tbl:libritts-mos}
	\vspace{-0.45cm}
\end{table*}
\begin{table}[t]
	\centering
	\caption{Pairwise subjective comparisons between BiVoC and representative baselines. Each cell reports the MOS difference and the Holm-adjusted $p$-value obtained using a two-sided Wilcoxon signed-rank test, where $\Delta\mathrm{MOS}=\mathrm{MOS}_{\mathrm{BiVoC}}-\mathrm{MOS}_{\mathrm{baseline}}$. Positive and negative values indicate higher and lower mean MOS for BiVoC, respectively.}
	\label{tbl:pairwise-mos-significance}
	\renewcommand{\arraystretch}{1.15}
	\setlength{\tabcolsep}{3.0pt}
	\resizebox{0.9\columnwidth}{!}{
		\begin{tabular}{@{}lcccc@{}}
			\toprule
			Dataset
			&HiFiGAN
			&Vocos
			& \shortstack{HiFiGAN-V1\\+BNN}
			& \shortstack{Vocos\\+BNN} \\
			\midrule
			AISHELL3
			&\shortstack{$+0.07$\\($p=0.260$)}
			&\shortstack{$-0.07$\\($p=0.260$)}
			&\shortstack{$+2.51$\\($p<0.001$)}
			&\shortstack{$+1.36$\\($p<0.001$)} \\
			MUSDB18
			&\shortstack{$+0.22$\\($p=0.002$)}
			&\shortstack{$+0.12$\\($p=0.027$)}
			&\shortstack{$+2.26$\\($p<0.001$)}
			&\shortstack{$+1.50$\\($p<0.001$)} \\
			\bottomrule
		\end{tabular}
	}
	\vspace{-0.15cm}
\end{table}

Fig.~{\ref{fig:mipr-miopr} presents the MIPR, MIOPR, MDPR, and MDOPR \emph{w.r.t.} different modules within the null-space module, where PESQ is chosen as the representative metric. Converting the head module, \emph{i.e.}, HSEM, from binary to full-precision yields the highest MIPR and MIOPR among all three modules. Conversely, binarizing the head from the full-precision baseline results in the highest MDPR and MDOPR. These consistent observations indicate that the head is particularly sensitive to binarization, \emph{i.e.}, retaining the head in full precision provides the largest PESQ improvement per unit additional resource, while binarizing it causes the largest degradation per unit resource saving.}

To further substantiate this observation, Table~{\ref{tbl:bnn-module-ablation}} presents the objective performance under different precision configurations. Starting from the fully binarized model, using a full-precision head notably improves performance with only minor increases in model size and computation. Although converting the body yields larger absolute improvements, it incurs a substantial increase in OPs, while converting the tail brings limited gains despite a large parameter increase. From the opposite direction, binarizing the head from the full-precision model causes pronounced performance degradation with only minor resource savings, whereas binarizing the body or tail provides a more favorable reduction in complexity. These complementary results demonstrate that retaining the head in full precision while binarizing the body and tail provides the most favorable quality-complexity trade-off, and this configuration is adopted in the subsequent study.


\textbf{Effect of RND strategy}: In Table~{\ref{tbl:rnd-ablation}}, we compare the objective performance in terms of whether to use RND strategy. Compared with the model without RND, the RND-based model achieves much lower M-STFT and MCD, as well as higher PESQ and GOMPSNR, demonstrating the importance of RND for BNN-based neural vocoder. Fig.~{\ref{fig:rnd-comparison}} further presents a qualitative comparison. Notably, without RND, the generated spectrum contains blurred harmonic structures and strong noise-like artifacts in the inter-harmonic regions. In contrast, with RND, these artifacts are well suppressed and the spectrum is closer to the reference. This fully indicates the effectiveness of the RND strategy in information preservation, which is significant for BNN-based vocoder and is adopted by default.

\textbf{Effect of subband strategy}: We explore the effectiveness of the adopted subband modeling strategy by comparing it with a full-band variant. Specifically, we replace the subband-based encoding and decoding modules with their full-band versions, and set the channel size to 1024 for a comparable modeling capacity. The resulting full-band variant contains 2.69 M trainable parameters and requires 1.29 Giga OPs per 5-second audio clip. As shown in Table~{\ref{tbl:subband-ablation}}, the subband-based model achieves better results across all objective metrics with fewer parameters, \emph{i.e.}, 1.08 M vs. 2.69 M, and comparable computational cost. These results indicate that full-band modeling is not well suited to BNN-based vocoding, since heterogeneous frequency regions are forced to be represented in a unified binary feature space. In contrast, subband modeling provides a finer representation granularity and allows the model to capture region-specific spectral characteristics more effectively, leading to a better quality-complexity trade-off. In Sec.~{\ref{sec:performance-effect-subband}}, we further compare the performance under different number of subbands.

\textbf{Comparisons of different modeling units}: We compare different basic modeling units under the binarized setting, as shown in Fig.~{\ref{fig:ablation-basic-units}} and Table~{\ref{tbl:units-ablation}}. Although directly binarizing existing powerful full-precision modules, such as ConvNeXtV2 and LKCAB, is a natural choice, we surprisingly observe notably worse performance than the proposed BCU. This suggests that modules designed for full-precision networks are not necessarily suitable for binarized design. Their complicated feature interaction mechanisms, such as feature modulation or gating-like operations, may amplify the representation error caused by 1-bit weights and activations, and thus hamper the modeling process.

By contrast, BCU adopts a simple residual feature flow, which provides a more stable information path for binary features and achieves the best objective results. In addition, removing normalization layers from BCU leads to consistent performance degradation, indicating that normalization is necessary to stabilize the numerical range of binary convolution outputs and maintain a balanced residual feature flow.

\textbf{Effect of normalization type}: In previous sections, LN is chosen by default. In this part, we compare different normalization strategies in Table~{\ref{tbl:norm-ablation}}. Unlike full-precision models, where the cost of normalization is usually negligible, normalization becomes a non-trivial component in BNN-based vocoders because binary convolution greatly reduces the convolutional OPs. According to our internal analysis, LN accounts for about 35.9\% of the overall OPs, motivating us to explore more efficient normalization designs.

Three normalization types are chosen: LN, batch normalization (BN), and DyT~{\cite{zhu2025transformers}}.  Among the three candidates, BN achieves the best scores while requiring the lowest OPs. This result suggests that BN provides a favorable quality-complexity trade-off for binarized vocoders. A possible reason is that binary operations lead to more discrete and locally fluctuating feature distributions, making sample-wise statistics used by LN less reliable. BN, instead, uses running statistics accumulated during training, which may provide a more global and stable normalization reference during inference. DyT reduces the computational cost compared with LN, but its lack of explicit mean-variance normalization may limit its ability to stabilize binary feature distributions. As a result, BN is selected by default in the subsequent studies.
\begin{figure*}[h]
	\centering
	\includegraphics[width=0.98\textwidth]{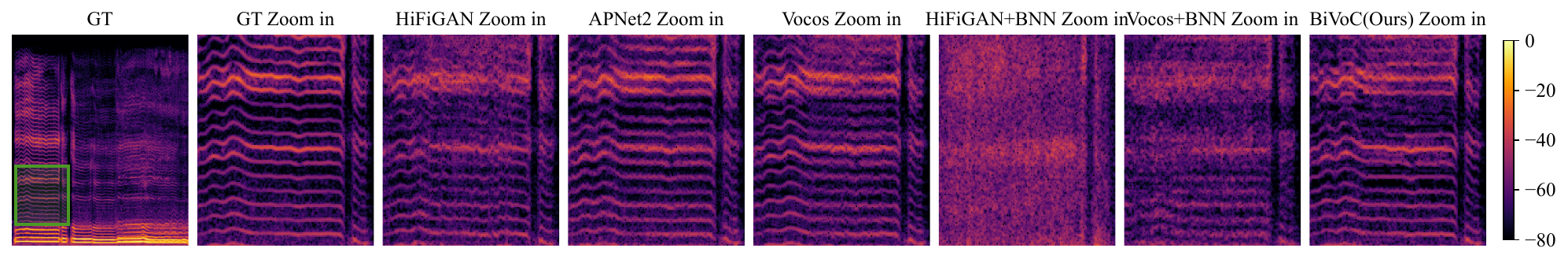}
	\caption{A typical visualization example of T-F spectra generated by different neural vocoders. From left to right: ground-truth (GT), HiFiGAN, APNet2, Vocos, HiFiGAN+BNN, Vocos+BNN, and the proposed BiVoC. The audio file is a vocal sound from the MUSDB18~{\cite{rafii2017musdb18}} test set.
	}
	\label{fig:spectral-visualization}
	\vspace{-0.40cm}
\end{figure*}
\begin{figure}[h]
	\centering
	\includegraphics[width=0.98\columnwidth]{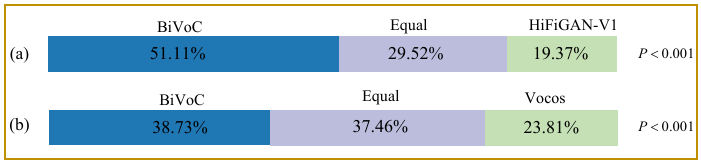}
	\caption{Average preference scores (in \%) of AB tests between two baselines and the proposed BiVoC. Audio clips are from RFWave test set.
	}
	\label{fig:ablation-abtest}
	\vspace{-0.45cm}
\end{figure}
\renewcommand\arraystretch{0.90}
\begin{table*}[t]
	\centering
	\caption{Objective performance under different number of subbands. Results are reported based on the LibriTTS test set.}
	\resizebox{0.98\textwidth}{!}{
		\begin{tabular}{cccccccccccc}
			\toprule
			\multirow{2}*{Ids} &Subband &\multirow{2}*{w/i Even} &\#Params. &OPs &\multirow{2}*{M-STFT$\downarrow$} &\multirow{2}*{PESQ$\uparrow$} &\multirow{2}*{MCD$\downarrow$} &V/UV$\uparrow$ &Period.$\downarrow$ &\multirow{2}*{GOMPSNR$\uparrow$} &\multirow{2}*{VISQOL$\uparrow$} \\
			&Number $F$ & &(M)  &(Giga/5s) & & & &F1 &RMSE  & & \\
			\midrule
			1 &7 &\xmark &\textbf{0.84} &\textbf{0.60}  &1.066 &2.916 &3.751 &0.922 &0.159 &4.033 &4.665 \\
			2 &13 &\xmark &0.91  &0.74  &1.021 &3.125 &3.538 &0.929 &0.151 &4.249 &4.716 \\
			3 &25 &\xmark &1.08 &1.01 &0.962 &3.447 &3.222 &0.938 &0.134 &4.645 &4.779 \\
			4 &25 &\cmark &1.05 &1.01  &0.994 &3.163 &3.498 &0.927 &0.153 &4.233 &4.728 \\
			5 &49 &\xmark &1.32 &1.57 &\textbf{0.937} &\textbf{3.528} &\textbf{3.127} &\textbf{0.939} &\textbf{0.130} &\textbf{4.853} &\textbf{4.806} \\
			\bottomrule
	\end{tabular}}
	\label{tbl:objective-different-subband}
	\vspace{-6pt}
\end{table*}
\renewcommand\arraystretch{1.0}
\begin{table}[t]
	\centering
	\caption{Different subband division schemes, including total number of subbands and bandwidths.}
	\begin{tabular}{cccc}
		\hline\hline
		Split-Type &Total Number &Subband Width (kHz) &Subband Number \\
		\hline
		\multirow{16}*{Uneven} &\multirow{4}*{7} &1 &3\\
		& &2 &2\\
		& &4 &1\\
		& &1 &1 \\
		\cline{2-4}
		&\multirow{4}*{13} &0.5 &6 \\
		& &1 &4 \\
		& &2 &2 \\
		& &1 &1\\
		\cline{2-4}
		&\multirow{3}*{25} &0.25 &12 \\
		& &0.5 &8\\
		& &1 &5\\
		\cline{2-4}
		&\multirow{4}*{49} &0.125 &24 \\
		& &0.25 &16 \\
		& &0.5 &8 \\
		& &1 &1\\
		\cline{2-4}
		\multirow{2}*{Even} &\multirow{2}*{25} &0.46875 &24 \\
		& &0.75 &1 \\
		\hline\hline
	\end{tabular}
	\label{tbl:subband-division-other}
	\vspace{-0.3cm}
\end{table}
\vspace{-0.25cm}
\subsection{Comparisons with Advanced Baselines}\label{sec:comparison-with-baselines}
\subsubsection{Objective comparisons} Table~{\ref{tbl:objective-metric-libritts}} reports the objective results on the in-distribution LibriTTS test set. Several valuable conclusions can be drawn. First, within the BNN category, the proposed BiVoC achieves the best performance across all metrics over two BNN-variants that are converted from full-precision baselines. This indicates that directly binarizing off-the-shelf neural vocoders is insufficient for high-fidelity waveform reconstruction. In contrast, the RND formulation and hierarchical subband modeling, together with the proposed BCU and mixed-precision design, effectively alleviate the information loss caused by binary representations. Second, compared with representative full-precision vocoders, the proposed BiVoC still shows competitive reconstruction quality under much lower theoretical model complexity. Although it does not surpass recent strong methods such as BigVGAN and DegVoC, it clearly outperforms several commonly used baselines like HiFiGAN and APNet2, and also approaches Vocos while using far fewer parameters and OPs. These results demonstrate that BiVoC provides a favorable quality-complexity trade-off for efficient audio generation. 

It is important to distinguish the theoretical equivalent OPs from the measured inference speed in Table~{\ref{tbl:objective-metric-libritts}}. To be specific, for the BNN-based models, the reported OPs are calculated following Eq.~{(\ref{eqn:40})}, assuming that binary operations are implemented using bit-packed XNOR-Popcount instructions. However, current PyTorch platform does not support such specialized binary kernels. The binarized activations and scaled binary weights are represented as floating-point tensors, and the subsequent convolutions are still executed using standard floating-point kernels. Moreover, the on-the-fly binarization and scaling operations introduce additional element-wise computation and memory-access overhead. Consequently, the theoretical reduction in equivalent OPs does not directly translate into proportional acceleration on the tested general-purpose CPU/GPU.

Further, one may ask with much fewer OPs, why is the BiVoC still slower than the full-precision Vocos and why is the BNN-based Vocos slower than the full-precision variant. For the first question, BiVoC performs hierarchical encoding and magnitude/phase decoding separately over $F=25$ subbands, which are executed sequentially. These operations lead to lower hardware utilization than the full-band and highly vectorized implementation of Vocos. For the second question, Vocos+BNN preserves a similar layer topology and tensor dimensions to Vocos while introducing additional on-the-fly binarization, but does not benefit from bit-packed binary convolution. Nonetheless, BiVoC still operates substantially faster than real time on both platforms. Developing bit-packed XNOR-Popcount kernels and parallelizing the subband branches will be investigated in future work.

Table~{\ref{tbl:ood-objective-comparisons}} presents the objective performance on another three out-of-distribution benchmarks, where models are pre-trained on the LibriTTS dataset. Again, the proposed BiVoC consistently outperforms the two BNN-based baselines by a large margin across the three test sets. BiVoC also surpasses several commonly used baselines like HiFiGAN and APNet2, and achieves competitive results with stronger T-F domain vocoders such as Vocos. These results demonstrate that the proposed method exhibits promising generalization capability on unseen audio domain. 

Fig.~{\ref{fig:spectral-visualization}} further provides a qualitative example from the out-of-distribution MUSDB18~{\cite{rafii2017musdb18}} test set. One can see that HiFiGAN+BNN and Vocos+BNN entirely fail to reconstruct harmonic structure. In contrast, BiVoC produces much cleaner harmonic trajectories and lower inter-harmonic noise floor, which is closer to the ground-truth spectrum. Compared with HiFiGAN, BiVoC also preserves more distinguishable harmonic details in the zoomed region, which further confirms the robustness of the proposed BNN-oriented vocoder.

\subsubsection{Subjective comparisons} MOS evaluations are conducted on the speech (AISHELL3) and music (MUSDB18) types, and a Friedman test is adopted to evaluate the overall statistical significance{\footnote{BigVGAN and DegVoC are included as strong full-precision references in the objective comparisons. The subjective tests focus on representative full-precision baselines to assess whether BiVoC can achieve competitive perceptual quality under a low-complexity binarized setting.
}}. As shown in Table~{\ref{tbl:libritts-mos}}, compared with three full-precision baselines, BiVoC obtains comparable MOS on AISHELL3 and the highest MOS on MUSDB18 vocals. The Friedman test results reveal significant overall differences on both AISHELL3 ($\chi_{F}^{2}\left(5\right)=90.32$, $p<0.001$) and MUSDB18 ($\chi_{F}^{2}\left(5\right)=95.48$, $p<0.001$), indicating that at least one vocoder differs from the others. To further determine the significance of the differences between BiVoC and representative baselines, we conduct post-hoc two-sided Wilcoxon signed-rank tests, with Holm correction for the four comparisons within each datasets. As shown in Table~{\ref{tbl:pairwise-mos-significance}}, on AISHELL3, BiVoC significantly outperforms both HiFiGAN+BNN and Vocos+BNN with statistical significance. No statistically significant difference is observed between BiVoC and either full-precision HiFiGAN or Vocos, although the mean MOS of BiVoC is slightly higher than that of HiFiGAN and slightly lower than that of Vocos. On MUSDB18, BiVoC achieves significantly higher MOS than all four baselines, with $p<0.05$ against HiFiGAN-V1 and Vocos, and $p<0.001$ against both BNN-based baselines. These results demonstrate that BiVoC substantially improves subjective quality over directly binarized vocoders, while maintaining competitive subjective quality relative to representative full-precision vocoders.

Fig.~{\ref{fig:ablation-abtest}} shows the AB preference scores between HiFiGAN/Vocos and the proposed BiVoC, audio samples are from the RFWave test set. The one-sided paired Wilcoxon test is conducted and the null-hypothesis is rejected if the $p$-value is lower than 0.05. Notably, the proposed method yields notable preference with statistical difference, further validating the effectiveness of our method. 
\vspace{-0.25cm}
\subsection{Performance of Different Subband Division}\label{sec:performance-effect-subband}
Recall that in Sec.~{\ref{sec:mis-configurations}}, the number of subbands $F$ is set to 25 by default. In this section, we investigate the effect of BiVoC under different $F$ as well as whether to split subbands in an uneven manner. As shown in Table~{\ref{tbl:objective-different-subband}}, four $F$ setups are considered: $\left\{7, 13, 25, 49\right\}$. For $F=25$, we also consider whether to split the subbands evenly. Detailed division configurations are illustrated in Table~{\ref{tbl:subband-division-other}}. Several conclusions can be drawn. First, the increase in number of subbands often leads to more fine-grained spectral modeling, which can be observed by the notable improvements in objective metrics, but also at the cost of increase in parameters and computational cost. $F=25$ seems to be a reasonable trade-off between performance and calculation complexity. Besides, uneven division strategy can result in substantial performance gains over the even scenario, as shown in Id3 and Id5, which suggests the unbalanced feature distribution of the spectrum.
\begin{table}[t]
	\centering
	\large
	\caption{Comparisons between three baselines and BiVoC under similar computational complexity. Results are reported based on the LibriTTS test set.}
	\label{tbl:bnn-ann-similar}
	\resizebox{0.92\linewidth}{!}{
		\begin{tabular}{c c c c c}
			\toprule
			\multirow{2}{*}{Models}  &OPs &\multirow{2}{*}{M-STFT$\downarrow$} &\multirow{2}{*}{PESQ$\uparrow$} &\multirow{2}{*}{GOMPSNR$\uparrow$} \\
			&(Giga/5s) & & & \\
			\midrule
			HiFiGAN-V1 &\textbf{0.99} &1.867 &1.829 &1.904 \\
			Vocos &0.98 &0.987 &3.196 &4.004 \\
			DegVoC &1.13 &0.965 &3.224  &4.305\\
			\midrule
			\textbf{BiVoC (Ours)} &1.01 &\textbf{0.962} &\textbf{3.447} &\textbf{4.645}\\
			\bottomrule
		\end{tabular}
	}
	\vspace{-0.0cm}
\end{table}
\begin{figure}[t]
	\centering
	\includegraphics[width=0.98\columnwidth]{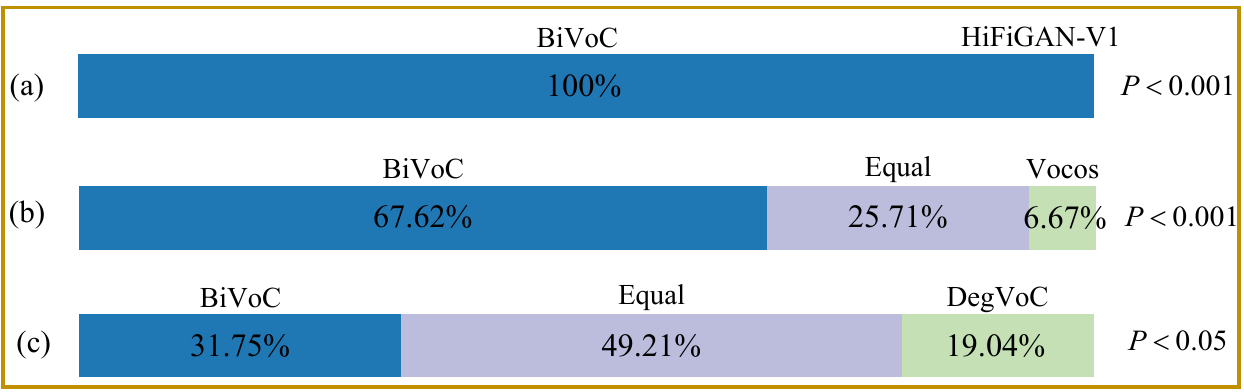}
	\caption{Average preference scores (in \%) of AB tests between the proposed BiVoC and three full-precision baselines under similar OPs. Audio clips are from RFWave test set.
	}
	\label{fig:bnn-ann-abtest}
	\vspace{-0.25cm}
\end{figure}
\vspace{-0.5cm}
\subsection{Comparisons with Full-precision Baselines with Similar Complexity}
There are several recent works that propose light-weight full-precision neural vocoders~{\cite{kong2020hifi,li2025learning}}. As such, a natural question arises: \textit{why not simply use a compact full-precision vocoder with similar computational cost?} To answer this question, we compare the proposed BiVoC with several advanced full-precision baselines whose OPs are adjusted to a similar scale.

Three baselines are chosen: HiFiGAN-V1, Vocos, and DegVoC. Their number of channels in each baseline are adjusted to exhibit similar OPs to BiVoC. Specifically, for HiFiGAN-V1, the number of input channels is squeezed to 40. For Vocos, the number of input channels is squeezed to 190, and the intermediate channel is set to 570. For DegVoC, the number of LKCABs is set to 4 and the number of channels is set to 32. As shown in Table~{\ref{tbl:bnn-ann-similar}}, the proposed method notably outperforms them in three objective metrics. In Fig.~{\ref{fig:bnn-ann-abtest}}, we also conduct the AB preference test between BiVoC and the three baselines. Notably, BiVoC is overwhelmingly preferred over HiFiGAN. When compared with Vocos and DegVoC, BiVoC also receives a clearly higher preference ratio with statistical significance. These results indicate that directly slimming down full-precision vocoders is not necessarily the optimal solution for low-complexity vocoder, and the proposed BNN-oriented design achieves a better quality-complexity trade-off. 
\section{Conclusion}\label{sec:conclusion}
In this paper, we propose BiVoC, a BNN-oriented T-F neural vocoder for high-fidelity waveform generation. Different from directly binarizing existing vocoder architectures, BiVoC is specifically designed according to the characteristics and limitations of binary operations. First, we leverage the range-null decomposition formulation to explicitly preserve the information contained in the mel spectrum, thereby reducing the reconstruction burden of the binarized generator. Second, we combine hierarchical subband modeling with a simple binarized convolutional unit specifically designed to improve representation granularity and feature modeling under 1-bit weights and activations. Third, based on a module-wise quality-complexity sensitivity analysis, we derive a mixed-precision configuration that better balances reconstruction quality and computational cost. Both quantitative and qualitative results demonstrate the effectiveness of the proposed method.

\lad{Limitations: Although our results demonstrate substantial theoretical reductions in computational and storage costs, the current PyTorch implementation does not exploit dedicated bitwise operations. Future work will therefore focus on hardware-aware optimization and real-device deployment with bit-packed binary kernels to further investigate the achievable improvements in inference latency, memory consumption, and energy on resource-constrained platforms.}


\bibliographystyle{IEEEtran}
\small
\bibliography{refs}

\end{document}